\documentstyle[amstex,prb,aps]{revtex}

\renewcommand{\vec}{\mathbf}
\renewcommand{\imath}{{\mathrm i}}
\renewcommand{\em}{{\it}}
\newcommand{\mat}{\mathbf}
\newcommand{\altmat}[1]{{\underline{\underline{\rm #1}}}}
\newcommand{\altvec}[1]{{\underline{\mathrm #1}}}
\newcommand{\op}{\mathcal}

\newcommand{\chemical}[1]{{$\rm #1$}}
\newcommand{\sign}{{\rm sign~}}
\renewcommand{\star}{{\rm star~}}
\renewcommand{\epsilon}{\varepsilon}
\newcommand{\bfsigma}{\mbox{\boldmath$\sigma$\unboldmath}}
\newcommand{\bfepsilon}{\mbox{\boldmath$\epsilon$\unboldmath}}
\renewcommand{\em}{\it}
\newcommand{\3}{\ss}

\begin{document}

\title{Models for the description of uniaxially modulated materials }
\author{BORIS NEUBERT, MICHEL PLEIMLING${}^\dagger$ and ROLF SIEMS}
\address{Theoretische Physik, Universit\"at des Saarlandes, Pf.\ 151150, D-66041 Saarbr\"ucken\\
${}^\dagger$present address:  Institut f\"ur Theoretische Physik B, Technische Hochschule,
D--52056 Aachen}

\maketitle

\begin{abstract}
Models which allow an  explicit application to structurally modulated
substances are reviewed within the frame of a symmetry-based
approach starting from discrete lattice theory. 
Focus is set on models formulated in terms of local variables assigned to
discrete crystallographic units (unit cells or parts of them).
Especially considered are symmetry-based pseudo spin models. Methods 
are discussed which permit the handling of the statistical mechanics of such
models, the translation of model calculation results to a form allowing the prediction
of experimental data, and the theoretical determination of point defect influences.\\
\end{abstract}
\vspace{0.5cm}

%
%
%
\bibliographystyle{unsrt}
%
%
%
%
%
%
%
%
\section{Introduction}
An interesting class of materials exhibits, as a distinguishing feature, phase diagrams
with several, in some cases with very many, commensurately and incommensurately
modulated phases. At present there exists a large body of experimental data on a
multitude of such materials, and various types of models have been proposed as bases
for a theoretical description. An assessment of their merits and their shortcomings
depends i) on the weight which one attaches to the physical justification of the general
form of the model and of the special values of its parameters, ii) on the properties of
the real system which are to be explained, and iii) on the degree of agreement between
theory and experiment which one requires.\par
At the present state of research it seems appropriate to survey a large class of these
models on the same footing, to discuss how they are connected with the original atomistic
or lattice theoretical basis, to find, if possible, a methodological and hierarchical
classification scheme into which they can be fitted, and to compare model predictions of
properties like phase
diagrams, values of physical quantities etc.\ with each other and with experimental data.\par
There exist extensive discussions of models exhibiting spatially
modulated structures\cite{Bli86a,Sel92}. In the present paper
it will be our special concern to discuss how different models are obtained when the basic first
principles theory is gradually more and more simplified so that finally theoretical
descriptions arise which allow explicit, though only approximate, calculations of physical
properties. We will be particularly interested in various models whose common feature is
that they describe the macroscopic specimens in terms of local variables assigned to 
discrete crystallographic units on the atomistic level. We will be especially concerned
with a class of such models, which arise if the continuous local variables are projected onto
a number of discrete states and, finally, with a description of these states in terms
of local pseudo spins with one or several components.\par
Besides the derivation of {\em model} characteristics -- e.g.\ the global structure of
phase diagrams, detailed information on the formation of new phases by structure
branchings and their accumulation points (which are connected with transitions from first
to second order phase transitions) or the exact structure of phase diagrams in the vicinity
of multiphase points -- we shall also report methods for establishing a connection
between theoretical (model) data and {\em experimentally accessible properties} of real systems.\par
An interesting additional aspect of the theoretical treatment is the possibility to 
incorporate impurities
into the theory. On the one hand, this line of the theory is of practical relevance for the
understanding of real, imperfect crystals; on the other hand it is of interest to find
out if the basic ideas of the modelling procedure carry so far as to 
allow also the description of impurity effects.\par
The paper is organized as follows. After a brief reference to the experimental
situation and some general remarks on the interrelation of model parameters and measured
quantities in section \ref{section:materials}, short comments on 
phenomenological theories of modulated structures and on the derivation of
pressure-temperature or temperature-electric field phase diagrams from Landau type
thermodynamic potentials are made in section \ref{section:phenomenology}. 
In section \ref{section:atomisticmodel}, a 
symmetry-based lattice theoretical description of 
uniaxially modulated structures is presented
which provides a framework for reviewing a number of models  in the subsequent sections
(models with continuous variables in section \ref{section:contimodels}, pseudo spin
models in section \ref{section:PSmodels}). Section \ref{section:application} describes
methods for bridging the gap between the results of model calculations (in terms of model
parameters) and experimentally observed properties (in terms of experimental control 
parameters).  These methods
do not postulate an ad hoc correspondence but -- with further approximations -- make 
use of the lattice theoretical foundation of the models and of macroscopic material
properties. The incorporation of point defects into the theory is discussed in
section \ref{section:pointdefects}. In section \ref{section:discoutlook}, 
a summarizing discussion and an outlook to future developments is given.\par
\section{Examples of modulated systems; objectives of their theoretical description}
\label{section:materials}
During the
late 70s and 80s a worldwide hunt\cite{Cum90,Saw92} for materials with cascades of structurally
modulated phases led to a large amount of experimental data on numerous compounds.  The
largest class of such crystals is formed by \chemical{A'A''BX_4}-compounds,
among which \chemical{Rb_2ZnCl_4} is a prototypical member. 
Apart from measurements under different temperatures, pressures and electric fields,
homologous series of \chemical{A'A''BX_4}-compounds, e.g.\ the 
tetramethylammonium tetrahalogenometallate (\chemical{TMA-MX_4}) series, allow the investigation of the
influence of the chemical composition on the sequence of phase transitions.\par
An outstanding specific material with an especially large number of commensurately or
incommensurately modulated phases is betaine calcium chloride dihydrate (BCCD). Its
phase diagram, which has been thoroughly investigated experimentally e.g.\ with respect
to its dependence on temperature, pressure, uniaxial stress, electric fields, 
or impurities shows many very interesting features. The observed global and local
details of the phase diagrams invite a comparison with theoretical predictions to test
the performance of various models.\par
We only briefly mention three further special examples 
since surveys on structurally modulated materials were given elsewhere 
(e.g.\cite{Cum90,Saw92,Per89,Bli86b}).
These examples are interesting in that they show that 
in families of isomorpheous compounds materials with properties deviating
significantly from the majority behavior may be observed:
a) All investigated modulated substances of the \chemical{TMA-MX_4} family
have their modulation wave vector $\vec q$
along ${\bf a}$ with the exception of \chemical{[N(CH_3)_4]_2CuBr_4}
with ${\bf q}$ parallel to ${\bf c}$ (similar to its homologue\cite{Sai91} 
te\-tra\-me\-thyl\-phos\-pho\-nium-\chemical{CuBr_4}). b) \chemical{Rb_2ZnCl_4}
and \chemical{Rb_2ZnBr_4} on the one hand and 
\chemical{Cs_2CdBr_4} and \chemical{Cs_2HgBr_4} on the other hand
show cascades of modulated phases but \chemical{Cs_2ZnCl_4} and \chemical{Cs_2ZnBr_4}
do not. c) Whereas most of the members of the \chemical{A_2BX_4} family exhibit
phase transitions of purely order-disorder type and hence lack 
a soft mode\cite{Axe86}, \chemical{K_2SeO_4} is primarily displacive 
in its incommensurate transition, but also possesses a significant order-disorder
component.\par
Most of the previously mentioned materials with uniaxial
structural modulations were investigated at ambient pressure over 
large temperature intervals; for some substances, pressure was also varied or
the influence of an electric field was studied.\par
Different approaches offer themselves for a theoretical interpretation of these data:
a rather rigorous treatment would start from the basic Hamiltonian formulated as a 
function of atomic positions, momenta, and externally imposed parameters 
(e.g.\ homogenous stresses and electric field components). This Hamiltonian might be 
deduced e.g.\ from an ab initio electron theory or from a formulation in terms of
empirical interatomic or intermolecular interactions. This treatment would then employ
the methods of statistical mechanics to derive for example thermodynamic potentials
for equilibrium structures and phase diagrams. These would be obtained as functions
of temperature and experimental control parameters and could be directly compared
with experimental results. If (due to computational difficulties) the method is not
(yet) feasible, one can make use of semiempirical treatments employing suitable
models. In the standard procedure for such a model description of modulated systems,
physical properties like phase diagrams are derived from effective model Hamiltonians
with the help of approximate methods of statistical mechanics. The results are
primarily obtained as functions of temperature and model parameters (including 
e.g.\ effective intercell interactions and electric field components).  For
a comparison with experimental data it would be desirable to find physically motivated
connections between the effective interactions occuring in the model Hamiltonian and
experimentally given parameters like temperature and applied stresses.\par
Most applications of model calculations to the interpretations of empirical data restrict
themselves, however, to a comparison of general topological features
of the model phase diagrams with those obtained experimentally or
try to establish an ad hoc correspondence.
In ref.\cite{Ten90b,Neu94b} an explicit mapping from a theoretical 
temperature-interaction-phase diagram to a pressure-temperature phase diagram
was presented which makes use of material properties like thermal expansion, elastic
constants etc. The method was used
exemplarily for BCCD and is described in subsection \ref{subsection:theoderivedpd}.\par
Phase diagrams were calculated e.g.\ for 
the DIFFOUR model\cite{Jan86a,Jan86b,Jan91},
the ANNNI model and its extensions\cite{Sel88,Yeo88}
the DIS model\cite{Ple94,Ple96}, 
the AANNDI model\cite{Kur94}, 
and for
Chen and Walker's model\cite{Che90,Che91b}.
Usually two-dimensional {\em planar}  parameter space sections were considered.\par
All models 
discussed in the sections \ref{section:atomisticmodel}--\ref{section:PSmodels}
contain variables that
correspond explicitely to microscopic crystalline quantities: both local generalized coordinates
or pseudo spins describe degrees of freedom of crystallographic (sub)cells. These model variables
are coupled by effective
interactions, which correspond to certain thermal averages of combinations of
original atomistic interactions. 
If it were possible to find a temperature-, stress-, and electric field-dependent
mapping of
the atomistic couplings onto the model parameters, precise predictions for
the occurence of phase transitions could be made. Particularly, it would be possible
for a special substance 
to transform a theoretical phase diagram.\par 
Because of their experimental relevance, phase diagrams as functions of temperature and
pressure are of special interest. The set of states experimentally accessible by variations
of $T$ and $p$ corresponds to a curved two-dimensional surface in the higher-dimensional
model parameter space spanned for example by temperature and effective interactions.
In general, this surface will not be planar.\par
%

%
\section{Phenomenological Theories}
\label{section:phenomenology}
As the phenomenological Landau theory has been the subject of various
reviews (see for example\cite{Tol87,Koc90}) we will focus in the following
on different schemes for describing successive transitions
between phases characterized by different wave numbers. 
We will especially discuss two approaches to
handle a cascade of commensurate phases.\par
In order to describe the cascade of phase transitions of a devil's
staircase Sannikov\cite{San89b,San90b,San91} proposed a
phenomenological theory in which the sequence of phase transitions is
determined by one soft branch of the lattice vibrations.
The thermodynamic potential for a transition to a commensurate phase characterized by the wave 
number $q_l=m/l$ can be written in the form
\begin{equation}
\Phi = \alpha \rho^2 + \beta \rho^4 - \alpha'_l \rho^{2l} \cos 2 l \zeta
\label{equation:eq1}
\end{equation}
where $\rho$ and $\zeta$ are the amplitude and the phase of the normal
coordinates. The coefficients $\alpha'_l$ differ from zero only
at the points of the Brillouin zone given by $q = q_l = \frac{m}{l}$
where $m$ and $l$ are integers, i.e.\ for every $l$ one obtains a 
different potential $\Phi$
(the wave numbers $q$ will be given in units of $2\pi/a_i,\,i=1,2,3$, $a_i$ being the lattice
 constants).
The limit $l \longrightarrow \infty$ corresponding to irrational 
values of $q$ yields the potential 
\begin{displaymath}
\Phi = \alpha \rho^2 + \beta \rho^4
\end{displaymath}
describing the transition to an incommensurate phase.\\
The coefficient $\alpha$ in eqn.\ (\ref{equation:eq1}) is a continuous
function of $q$. The approximation used for this function depends upon
the interval of $q$-values considered.
If the end points of the interval do not satisfy the Lifshitz condition,
$\alpha$ is approximated by the
expression\cite{San89b}
\begin{displaymath}
\alpha = a + \delta \left( q - b \right)^2.
\end{displaymath}
Here $b$ is the wave number of the soft mode responsible for the minimum
of the branch $\alpha (q)$. 
This first case is realized for example in the \chemical{TMA-MCl_4} family.
If, on the contrary, one of the end points
satifies the Lifshitz condition (e.g.\ $q = 0$) the proposed form 
is\cite{San90b}
\begin{displaymath}
\alpha = A + \kappa \left( q^2 - B^2 \right)^2
\end{displaymath}
which results from an expansion in $q^2$ around the point $q = 0$. 
Examples for this case are thiourea [\chemical{SC(NH_2)_2}]
and BCCD\cite{San91}. In the following we will focus on the latter case.\par
Introducing new dimensionless variables $x = \frac{B}{Q}$ and
$y = \frac{A}{\kappa Q^4}$ and assuming that the value $|\alpha'_l|$
is the same for all $l$ Sannikov derived a phase diagram in the
$x$-$y$-space. $Q$ has the same dimension as $q$ and its value is
arbitrary. Under the assumption of a linear $T$-dependence of
$A$ and $B$ (and therefore of $x$ and $y$) the phase 
sequences were obtained for different materials.\par
In a detailed analysis for BCCD, thermodynamic 
potentials for various values of $q$ were derived.
Using an ad hoc linear transformation from dimensionless model variables to
$T$ and $p$ and choosing tentative values for the remaining parameters
a rather satisfying fit of theoretical to experimental
$T$-$p$ phase diagrams could be obtained\cite{San97}.\par
In order to study the influence of external (electric) fields $E_i$
coupling terms between the fields and the two-component order parameter
were added to the potential (\ref{equation:eq1}). The transformation
properties of the conjugated quantities $P_i = - \frac{\partial \Phi}
{\partial E_i}$ (spontaneous polarization)
could than be derived and compared to experiment\cite{San91}.\par
A different phenomenological theory was developed especially for
the interpretation of the phase transition sequence of
BCCD\cite{Rib90,Alm92,Cha93}. Noticing
that the polarization of the ferroelectric low temperature phase
($q = 0$) corresponds to the primary order parameter, the starting point
for this approach is the free energy density (in reduced units)
\begin{displaymath}
g(x) = t \xi^2 + \frac{1}{4} \xi^4 - \frac{1}{4} \left( \frac{
\partial \xi}{\partial x} \right)^2 + \frac{1}{4} \left( \frac{
\partial^2 \xi}{\partial x^2} \right)^2 + \mu \xi^2 \left(
\frac{ \partial \xi}{\partial x} \right)^2.
\end{displaymath}
The order parameter is supposed to be site-dependent. The terms
$- \frac{1}{4} \left( \frac{ \partial \xi}{\partial x} \right)^2$
and $\frac{1}{4} \left( \frac{ \partial^2 \xi}{\partial x^2} \right)^2$
describe the occurence of modulations whereas the term
$\mu \xi^2 \left( \frac{ \partial \xi}{\partial x} \right)^2$
is responsible for the temperature dependence of the wave 
vector\cite{Rib90}. In addition umklapp terms were introduced into the
expression for the free energy density
of the commensurate phases and approximated by a forth degree term
in the order parameter amplitude,
\begin{displaymath}
U_{\text{umklapp}} = - b_{2l} \xi^{2l} \approx - \frac{1}{4} \beta_{\text{eff}} \left( q
\right) \, \xi^4,
\end{displaymath}
for a commensurate phase with wave number $q = \frac{m}{l}$.\\
Using the plane-wave approximation (i.e.\ $\xi = \xi_0 \cos k x$)
and choosing appropriate values for
$\beta_{\text{eff}} \left( q \right)$, 
the phase diagram was derived by computing the free energies of the different phases
as functions of temperature\cite{Rib90}. 
Introducing coupling terms
between the order parameter and an electric field\cite{Cha93}
the $E-T$ phase diagram was calculated. In a subsequent paper\cite{Alm92}
the sinusoidal scheme was abandoned and the occurence of a multisoliton
regime was considered.

%
\section{Derivation of models from an atomistic lattice theoretical basis}
\label{section:atomisticmodel}
%
%
\subsection{Formulation in terms of local modes}

Models based on the discrete structure of the lattice can
provide a way to achieve a physical understanding of the processes on the atomic level
of the crystal which goes beyond a mere phenomenological description. Such microscopic
models can be handled analytically or numerically. In the subsequent sections models which have been closely
studied with respect to the application to experimentally investigated substances
are discussed. Each model
variable -- either continuous as in the case of the DIFFOUR models or discrete as in the
case of pseudo spin models -- usually describes one degree of freedom per lattice site.  
Previously it was considered a weak spot, especially of pseudo spin models, 
that the significance (on
an atomistic scale) of the model (pseudo spin) variables was not well defined. For this reason
we summarize a general procedure for the systematic definition of  symmetry-based 
model variables on a lattice theoretical basis.
This procedure stresses the atomistic roots of the models, leads to models
conforming to lattice symmetry and facilitates the prediction of spontaneous
polarizations of modulated phases 
and of the atomic displacements occuring at structural transitions\cite{Neu97}.
It combines the idea of local modes formulated e.g.\ in ref.\cite{Tho71} 
with the rigorous symmetry considerations of Landau's theory (for instance, 
see ref.\cite{Tol87,Koc90}). It leads to a Hamiltonian that serves as a starting
point for various approximations leading to different models (e.g. the DIFFOUR models or
the DIS model) and thus provides a frame for reviewing very different models from a
common point of view.\par
The key idea of the general procedure for the derivation of a microscopic model Hamiltonian
of this procedure is
the introduction of variables as generalized coordinates of symmetry-adapted local modes 
(SALMs). This facilitates the incorporation of both the discreteness of the
lattice and the overall symmetry of the crystal. The symmetry properties of the
SALMs can either be derived from first-principles calculations or taken from
experimental data, i.e.\ from the observation of the symmetry of the displacements
responsible for the phase transitions.\par
Let the crystal be composed of $N$ unit cells each containing $K$
particles (of which $K'$ are non equivalent) and assume, for
definiteness, Born-von K\'arm\'an periodic boundary conditions. The crystal structure is  then
described by a $3KN$-dimensional configuration vector $\altvec R$, where each entry specifies
a component of the position of an atom. An arbitrary configuration 
\begin{displaymath}
{\altvec R}= {\altvec R}_0 + \delta{\altvec R}
\end{displaymath}
is decomposed into a contribution ${\altvec R}_0$, which is invariant under all elements
${\op G}=\{{\mat R}|{\vec t}\}$ of the high temperature normal phase space group $\frak G_0$,
and a symmetry-breaking displacement $\delta{\altvec R}$, which is zero above the transition 
temperature $T_{\text{crit}}$ 
from the normal phase to the lower symmetry 
phases (unmodulated or commensurately or incommensurately modulated).  
The displacement $\delta{\altvec R}$ from a reference structure
at given reference values of $T$, $p$, etc.\ can be written as a superposition 
\begin{equation}
\delta{\altvec R}= \sum_{\vec n} \delta{\altvec R}_{\vec n}
\label{equation:Wexpansion}
\end{equation}
of local contributions $\delta{\altvec R}_{\vec n}$. They describe displacements for only
those atoms which are associated to unit $\vec n$, having zero entries for all others.
These units do not necessarily correspond to the unit cells of the crystal; in
the case of the space group $Pnma$ discussed in subsection \ref{subsection:egPnma}, 
half cells are chosen
as units accounting for the pseudoperiodicity of such crystals along one 
of the lattice vectors.
$\delta{\altvec R}_{\vec n}$ is expanded in terms of local modes. In most cases it is
sufficient to retain only a few (e.g.\ one or two) local modes per unit.\par
It is useful to choose the local modes such that they reflect the
crystal symmetries:
the smallest unit in the crystal with respect to symmetry 
is the asymmetric unit ${\frak A}^1_{\vec 0}$ of the space group $\frak G_0$.
Symmetry imposes no restrictions on the position of the $K'$ atoms associated
to this polyhedral subcell. Further
subcells ${\frak A}^i_{\vec n}$ defined by
\begin{displaymath}
{\frak A}^i_{\vec 0}= {\op G}^i {\frak A}^1_{\vec 0} \mbox{ and } {\frak A}^i_{\vec n}=\{{\mat E}|{\vec n}\} {\frak A}^i_{\vec 0}
\end{displaymath}
cover the whole crystal. The
$B$ space group operations ${\op G}^1,\ldots,{\op G}^B$ are coset representatives 
in a coset decomposition of $\frak G_0$ with respect to its invariant 
subgroup $\frak T$ of primitive
translations. The index $B=[{\frak G_0}:{\frak T}]$ equals the order of the point group
$\widehat{\frak G}_0$ of ${\frak G_0}$ and determines the number of subcells ${\frak A}^i_{\vec n}$
per cell $\vec n$. The coset representatives have to be chosen such that these $B$ subcells
form a contiguous new choice of cell. Any displacement of the $K'$ atoms associated with
${\frak A}^i_{\vec n}$ can be expanded in terms of a basis set of $3K'$ subcell modes 
${\altvec V}^{i\kappa}_{\vec n}, \kappa=1,\ldots,3K'$. They should satisfy the relation
${\altvec V}^{i'\kappa}_{\vec n'}={\op G}{\altvec V}^{i\kappa}_{\vec n}$ where
${\op G}\in{\frak G_0}$ is the space group operation transforming subcell 
${\frak A}^{i}_{\vec n}$ into ${\frak A}^{i'}_{\vec n'}$. Thus, the modes in any subcell
are entirely determined by the (arbitrary) choice of modes in the
asymmetric unit ${\frak A}^1_{\vec 0}$.\par
A complete -- for $K'B>K$ over-complete (see below) -- set of $3K'BN$ local 
modes can be constructed as linear combinations of the subcell modes 
\begin{equation}
\label{equation:WSV}
{\altvec W}^{I\kappa}_{\vec n}=\sum_{i=1}^B S^{Ii} {\altvec V}^{i\kappa}_{\vec n}
\end{equation}
with a $B \times B$ transformation matrix $\altmat S$, the form of which is
determined by
the IRREPs of the point group $\widehat{\frak G}_0$.
For fixed $I\kappa$ each local mode
${\altvec W}^{I\kappa}_{\vec n}$ describes the {\em same} displacement pattern of atoms 
in their respective unit, 
i.e.\ ${\altvec W}^{I\kappa}_{{\vec m}+{\vec n}}=\{{\mat E}|{\vec m}\} {\altvec W}^{I\kappa}_{\vec n}$.
We will call the set $\{{\altvec W}^{I\kappa}_{\vec n}: \kappa=1,\ldots,3K'\}$ the
$I$th set of symmetry-adapted local modes for unit $\vec n$. Any linear combination
of displacement vectors in one set of SALMs has the same symmetry properties.
Apart from a primitive translation, an 
arbitrary space group operation only permutes the subcell indices and thus 
the transformation behavior of the set of ${\altvec W}^{I\kappa}_{\vec n}$ is entirely
determined by $\altmat S$. The actual positions of the atoms in the subcells have not
to be known.
Only those sets of SALMs can contribute to the structural deformation $\altvec U$ that
have the symmetry properties imposed by the IRREP according to which $\altvec U$  transforms.\par
The existence of atoms on special positions, i.e.\ on faces, edges and corners of the
subcells causes the number $3K'BN$ of local modes to be in general higher than the number
$3KN$ of independent displacements. Hence, the coefficients of an expansion of a general
displacement in terms of these SALMs have to satisfy consistency relations.\par
The symmetry-breaking displacement $\delta{\altvec R}_{\vec n}$  in unit $\vec n$
is decomposed 
into contributions $\delta{\altvec R}^I_{\vec n}$ from the $B$ sets of SALMs:
\begin{equation}
\label{equation:deltaRn}
\delta {\altvec R}_{\vec n}= \sum_{I=1}^B \delta {\altvec R}_{\vec n}^I= \sum_{I=1}^B \sum_{\kappa=1}^{3K'} a_{\vec n}^{I\kappa} {\altvec W}_{\vec n}^{I\kappa}.
\end{equation}
Under a continuous change of external parameters the average positions of the 
atoms are displaced in such a way 
that a valley of
low energy is followed in the high-dimensional configuration space. 
The symmetry-breaking
displacement of any atom in the asymmetric unit ${\frak A}^1_{\vec 0}$ which is 
connected with the transition is not
determined by space group symmetry. It will, in general,  be {\em different} even for materials
exhibiting the {\em same} symmetries of the normal and the modulated phases 
(e.g.\ \chemical{A_2BX_4} compounds). 
Hence, the contribution $\delta{\altvec R}_{\vec n}^I = \sum_{\kappa=1}^{3K'} a_{\vec n}^{I\kappa} {\altvec W}_{\vec n}^{I\kappa}$
from the $I$th set of SALMs will, in general, lie on a $d_a\leq3K'$-dimensional 
curved surface 
or, in its simplest and most frequent case, describe a curved path ($d_a=1$) upon changing
temperature or pressure. Either case corresponds to a simultaneous variation of 
in general {\em all}
coefficients $a_{\vec n}^{I\kappa}$ in  equation (\ref{equation:deltaRn}) as the fixed
subcell modes ${\altvec V}^{i\kappa}_{\vec n}$ can not simultaneously be adapted to the
curved paths described by every $\delta{\altvec R}_{\vec n}^I$.
Thus, instead of using the coefficients from 
equation (\ref{equation:deltaRn})
as generalized variables for the determination of the displacements in unit $\vec n$,
one introduces new local sets of curvilinear coordinates for each set of SALMs
by means of a coordinate transformation 
$a_{\vec n}^{I\kappa}= f^{I\kappa}(Q^{I1}_{\vec n},\ldots,Q^{I,3K'}_{\vec n}),\,
\kappa=1,\ldots,3K'$ such that
the contribution 
\begin{displaymath}
\delta{\altvec R}_{\vec n}^I(Q^{I1}_{\vec n},\ldots,Q^{I,3K'}_{\vec n})=
\sum_{\kappa=1}^{3K'}f^{I\kappa}(Q^{I1}_{\vec n},\ldots,Q^{I,3K'}_{\vec n}){\altvec W}_{\vec n}^{I\kappa}
\end{displaymath}
to the symmetry-breaking displacement in unit 
$\vec n$ [cf.\ equation (\ref{equation:deltaRn})]
occurs when $Q^{I1}_{\vec n},\ldots,Q^{Id_a}_{\vec n}$ increase from
zero to some non-zero value whereas all other generalized variables remain at their
thermal average of zero. In the simplest case $d_a=1$, only one relevant 
variable $Q^{I1}_{\vec n}$ per set of SALMs is needed and thus the value 
of $Q^{I1}_{\vec n}$ determines a linear combination
of displacement vectors from the $I$th set of SALMs for unit ${\vec n}$ (which
has the same transformation behavior as any other linear combination of SALMs
from the same set).\par
On the one hand, the space group symmetries and the direction of the modulation determine the
subcell generators ${\op G}^1,\ldots,{\op G}^B$ and the matrix $\altmat S$ 
and thus should be the same
for a whole class of materials. On the other hand, the displacements of atoms in the
asymmetric unit and thus the coordinate transformations $f^{I\kappa}$ 
are determined by the local potential and thus by the specific material under investigation.
This allows a separation of properties specific for a certain class of materials from
properties specific for the substance under investigation.\par 
The vector field $\delta{\altvec R}_{\vec n}^I(Q^{I1}_{\vec n},\ldots)$ 
gives a contribution to the structural transformation in unit $\vec n$ 
with definite transformation behavior. In general, $d_s$ sets of SALMs ($I=1,\ldots,d_s$
without loss of generality) are needed for
the construction of symmetry modes 
transforming according to IRREPs of the group $\frak G({\vec q})$ of the wave vector.
When $\widehat{\frak G}_0$ has only one-dimensional real 
IRREPs and only {\em one}
order parameter transforming according to one IRREP of
$\frak G({\vec q})$ gives the primary contribution to the structural deformation 
(as it is the case in BCCD), only one ($d_s=1$) certain set of SALMs 
($I=1$) and the corresponding symmetry-breaking displacement 
$\delta {\altvec R}_{\vec n}^1$ is needed for every unit. 
The reverse is not true because by different spatial modulations of $Q^{11}_{\vec n}$ one set
of SALMs can be used to construct symmetry modes transforming according to different
IRREPs of ${\frak G}({\vec q})$ and different sets of SALMs may contribute to
symmetry modes with same transformation behavior (e.g.\ at a point ${\vec q}\not={\vec 0}$ 
inside the
Brillouin zone when ${\frak G}({\vec q})$ is a true subgroup of ${\frak G}_0$; 
examples will be given in 
subsection \ref{subsection:egPnma}). Especially, the set of SALMs
corresponding to the totally symmetric representation of $\widehat{\frak G}_0$
may be relevant because it may contribute to a structural deformation not transforming
according to the totally symmetric representation of ${\frak G}_0$.\par
In the simplest case of one relevant set of SALMs ($d_s=1$) and one degree of freedom in
the asymmetric unit ($d_a=1$), expansion (\ref{equation:Wexpansion}) can be written
\begin{displaymath}
\delta{\altvec R}= \sum_{\vec n} \delta{\altvec R}^1_{\vec n}(Q^{11}_{\vec n},0,\ldots,0)
\end{displaymath}
and the $3KN$-dimensional lattice theoretical problem is reduced to a system with
one generalized local variable $Q^{11}_{\vec n}$ per unit.
In the next paragraph,  for simplicity, we will restrict ourselves to this special case,
which however can easily be generalized.\par
The crystal potential is expanded in terms of the local generalized coordinates 
$Q^{I\kappa}_{\vec n}$. The
terms containing only variables with given $\vec n$ form
the local potential $\Phi^{\text{loc}}(Q^{11}_{\vec n},\ldots,Q^{B,3K'}_{\vec n})$ for
unit $\vec n$. The general Hamiltonian of the system can then be written
\begin{displaymath}
{\op H}=\sum_{\vec n} \left[ \sum_{I,\kappa} \left(P^{I\kappa}_{\vec n}\right)^2 + 
\Phi^{\text{loc}}(Q^{11}_{\vec n},\ldots,Q^{B,3K'}_{\vec n})\right] +
\Phi^{\text{int}}(\ldots,Q^{I\kappa}_{\vec n},\ldots,Q^{J\lambda}_{\vec m},\ldots)
\end{displaymath}
where $P^{I\kappa}_{\vec n}$ is a momentum canonically conjugate to $Q^{I\kappa}_{\vec n}$.
Averaging out all irrelevant variables  yields an effective Hamiltonian
\begin{equation}
\label{equation:Hamil1}
{\overline{\op H}}= {\rm tr}(\rho {\op H}) = \sum_{\vec n} \left[\left(P^{11}_{\vec n}\right)^2 + {\overline{\Phi}}^{\text{loc}}(Q^{11}_{\vec n})\right] + {\overline{\Phi}}^{\text{int}}(\ldots,Q^{11}_{\vec n},\ldots,Q^{11}_{\vec m},\ldots).
\end{equation}
(The trace is taken only for the variables to be averaged out: $I\not=1,\kappa\not=1$.)\par
In the case of two relevant sets of SALMs per unit ($d_s=2,\,I=1,2$), 
one gets the two-variable version
\begin{eqnarray*}
{\overline{\op H}}&=& \sum_{\vec n} \left[\left(P^{11}_{\vec n}\right)^2 + \left(P^{21}_{\vec n}\right)^2 + {\overline{\Phi}}^{\text{loc}}(Q^{11}_{\vec n},Q^{21}_{\vec n})\right] \\
&&\qquad\qquad+ {\overline{\Phi}}^{\text{int}}(\ldots,Q^{11}_{\vec n},Q^{21}_{\vec n},\ldots,Q^{11}_{\vec m},Q^{21}_{\vec m},\ldots).
\end{eqnarray*}
The same form with the superscript $(21)$ replaced by $(12)$ is obtained in the case 
of one relevant set of SALMs per unit ($d_s=1,\,I=1$) but two relevant generalized 
coordinates ($d_a=2$) per asymmetric subcell. 
We will omit the superscript referring to $\kappa=1$ for brevity in the case $d_a=1$ 
of one generalized variable for each set of SALMs.

\subsection{Relation to other models}

The models discussed in the subsequent sections can be viewed as special cases of
the Hamiltonian (\ref{equation:Hamil1}) or its two-variable versions. The total number
of relevant variables per unit is $d_a\cdot d_s$. The local potential
${\overline{\Phi}}^{\text{loc}}$ and the interaction potential ${\overline{\Phi}}^{\text{int}}$ 
are effective potentials which, by their very definition, depend directly on elastic 
stresses (which determine the lattice constants) and on temperature due to
the thermal averaging over the irrelevant variables.
The potentials are usually
expanded up to forth or second order respectively and only a subset of interactions between 
nearest, next nearest and third next nearest neighbors is kept.\par
As in Landau theory, space group operations can be considered to act on the generalized
variables instead on the modes ${\altvec W}^I_{\vec n}$. Hence, symmetry 
considerations give 
conditions for the
dependence of the potential on the generalized variables, which have for example
been taken into account
when formulating the AANNDI or the DIS model. Moreover, the local potential may have 
equivalent minima: assume that there is one global minimum of the local potential for a
certain configurational set of generalized local coordinates; then
every space group operation which transforms a unit into itself 
but changes the configuration of the
respective coordinates may generate another equivalent minimum.\par
A further simplification may be introduced by projecting the
remaining continuous generalized local variables onto two-valued
pseudo spin variables as the signs of the former. This leads
to different pseudo spin models: we discussed either the ANNNI model 
in the case of one relevant mode per half cell or
the DIS model in the case of two relevant variables 
earlier\cite{Neu94a,Neu96}.\par
The Hamiltonian (\ref{equation:Hamil1}) corresponds to a version of the DIFFOUR model 
with one degree of freedom per lattice site or to the ANNNI model.
The more general case where two local modes with different transformation behavior 
($d_s=2$) are
incorporated corresponds to Chen and Walker's model or to the DIS model. Assuming two 
relevant displacement vector fields for the asymmetric unit ($d_a=2$), i.e.\ two generalized 
variables $Q^{11}_{\vec n}$ and $Q^{12}_{\vec n}$ for the relevant set of SALMs,
one arrives at a model
corresponding to the version of the DIFFOUR model with two degrees of freedom per lattice
site discussed in ref.\cite{Hli96,Qui96} ($Q^{11}_{\vec n}$ and $Q^{12}_{\vec n}$ 
correspond to variables describing the displacements of the betaine and calcium chloride 
groups of BCCD respectively). For the description of BCCD with
space group ${\frak G}_0=Pnma$, the units correspond to {\em half cells}. The latter two cases
yield  Hamiltonians with two different variables for each of such {\em units}.\par
It is noteworthy that the Hamiltonian (\ref{equation:Hamil1}) contains only one 
symmetry-breaking contribution per unit. As discussed above, the totally symmetric IRREP
of ${\frak G}_0$, which also
contributes to the deviation of the displacements from the structure of the para phase at the
original reference values of $T$, $p$, etc., is always present but has been absorbed into the atom 
positions of a new reference state.
The latter can be viewed as giving a background contribution to the structural
deformation of the crystal which is present throughout the whole range of temperature and
pressure.\par
%

%

\subsection{Exemplary study of the procedure for materials with space group $Pnma$}
\label{subsection:egPnma}

Among the  66 crystals with structurally incommensurate phases listed by Cummins\cite{Cum90},
the 22 materials belonging to the space group ${\frak G}_0=D_{2h}^{16}$ (in
their normal phase) form
the largest class. The materials  in the \chemical{A_2BX_4} (\chemical{K_2SeO_4}) family
(apart from a small group of tungstates and molybdates), betaine calcium chloride
dihydrate (BCCD), ammonium oxalate hemihydrate (AHO) and thiourea   all
exhibit the symmetries described by that space group.
Depending on the particular choice of axes, various Hermann-Mauguin
symbols have been attributed to these substances.
For a more detailed discussion see  ref.\cite{Cum90}.
In the present study, we choose the assignment $Pnma$ und restrict ourselves to
modulations along $\vec c$.
Special attention is payed to the pseudoperiodicity. The reader should not get confused
by the formalism as in the present case 
units correspond to {\em half cells} but the labeling with
{\em cell} subscripts ${\vec n}$ is kept for simplicity.\par
The asymmetric unit ${\frak A}_{\vec 0}^1=[0;\frac{1}{2}] \times [0 ; \frac{1}{4}] \times [0 ; 1]$
of the space group $Pnma$ is a parallelepiped (,block') 
containing all $K'$ nonequivalent
atoms from which the whole crystal can be filled by
the $B=8$ space group operations 
\begin{alignat*}{4}
{\op G}^1  &=  \{ {\mat E} | 000 \},\,& 
{\op G}^2  &=  \{ \sigma_y | 0 \frac{1}{2} 0 \},\,&
{\op G}^3  &=  \{ {\mat C}_2^y | 1 \frac{1}{2} 1 \},\,&
{\op G}^4  &=  \{ {\mat I} | 1 1 1\}\\
{\op G}^5  &=  \{ {\mat C}_2^z | \frac{1}{2} 1 \frac{1}{2} \},\,&
{\op G}^6  &=  \{ \sigma_x | \frac{1}{2} \frac{1}{2} \frac{1}{2} \},\,&
{\op G}^7  &=  \{ {\mat C}_2^x | -\frac{1}{2} \frac{1}{2} \frac{3}{2} \},\,&
{\op G}^8  &=  \{ \sigma_z | -\frac{1}{2} 0 \frac{3}{2} \}
\end{alignat*}
and primitive translations $\{{\mat E}|n_1 {\vec e}_1 + n_2 {\vec e}_2 + n_3 {\vec e}_3\}$.
The eight blocks (see figure \ref{figure:blocks}) generated from the asymmetric unit by the operations
${\op G}^1,\ldots,{\op G}^8$ are considered to form a new choice of cell
at ${\vec n}=(000)$. An arbitrary block is labeled by the index $i$
of the space group operation and the vector ${\vec n}$ 
of the primitive translation it was generated by. Blocks $1,\ldots,4$
extend from $z=n_3$ to $n_3+1$ (first half of cell),
blocks $5,\ldots,8$ from
$z=n_3+1/2$ to $n_3+3/2$ (second half of cell) along the $\vec c$-axis.\par
Local modes ${\altvec W}_{\vec n}^{I\kappa}$ are constructed from
the subcell displacement vectors ${\altvec V}_{\vec n}^{i\kappa}$ by means of
the $8 \times 8$-matrix (cf.\ eqn.\  \ref{equation:WSV})
\begin{displaymath}
{\altmat S}=\frac{1}{2}\left(\begin{array}{rrrrrrrr}
 1& 1& 1& 1& 0& 0& 0& 0\\
 1& 1&-1&-1& 0& 0& 0& 0\\
 1&-1& 1&-1& 0& 0& 0& 0\\
 1&-1&-1& 1& 0& 0& 0& 0\\
 0& 0& 0& 0& 1& 1& 1& 1\\
 0& 0& 0& 0& 1& 1&-1&-1\\
 0& 0& 0& 0& 1&-1& 1&-1\\
 0& 0& 0& 0& 1&-1&-1& 1\end{array}\right).
 \end{displaymath}
Although constructed for the whole (new) unit cell, these local modes describe
displacements for either the first ($I=1,2,3,4$) or the second half ($I=5,6,7,8$) of the
cell when viewed along the $\vec c$-direction in accordance with the definition of the blocks.
Thus half cells (quadruples of blocks) 
are the units used to introduce the local potential. Nevertheless, we do not relabel
the half cells but distinguish between front and back half cells by means of the
subscript $\vec n$ {\em and} the superscript $I$.\par
In Landau theory,
$\delta{\altvec R}$ is decomposed into symmetry modes ${\altvec e}^{{\vec k}\alpha\lambda}_s$
of the crystal transforming 
according to irreducible representations (IRREPs) $\Gamma^{{\vec k}\alpha}$ of ${\frak G}_0$:
\begin{equation}
\label{equation:LandauExp}
\delta{\altvec R}= \sum_{{\star}{\vec k}} \sum_\alpha \sum_\lambda \sum_{s=1}^{\dim \Gamma^{{\vec k}\alpha}} c^{{\vec k}\alpha\lambda}_s {\altvec e}^{{\vec k}\alpha\lambda}_s,
\end{equation}
where the first sum runs over all distinct stars, $\alpha$ labels the different IRREPs 
connected with $\star{\vec k}$, $\lambda$ accounts
for the fact that the IRREP $\Gamma^{{\vec k}\alpha}$ can appear more than once and
the sum over $s$ extends over all basis function of the IRREP $\Gamma^{{\vec k}\alpha}$. 
Contributions from the totally symmetric IRREP of ${\frak G}_0$ 
are included in ${\altvec R}_0$ and  excluded in equation (\ref{equation:LandauExp}) 
by the definition of $\delta{\altvec R}$. Below $T_{\text{crit}}$  at least one IRREP
gives a non-zero contribution to equation (\ref{equation:LandauExp}) and the respective 
symmetry-breaking displacement determines the structure of the newly formed phase.
In the microscopic theory presented in this section,
the symmetry modes $c^{{\vec k}\alpha\lambda}_s {\altvec e}^{{\vec k}\alpha\lambda}_s$
are formed by superposition of the local mode vector fields
$\delta{\altvec R}^I_{\vec n}$. The generalized variables $Q^{I1}_{\vec n},\ldots$ 
are spatially modulated order parameters.\par
It follows from the transformation behavior of the local modes under arbitrary elements
of the space group ${\frak G}_0$ that it is possible to construct symmetry modes 
transforming according to the IRREPs $\Lambda_2$ or $\Lambda_3$ 
(but not $\Lambda_1$ or $\Lambda_4$) if one takes into
account local modes with superscripts $I=3,4,7,8$ and transforming 
according to $\Lambda_1$ or $\Lambda_4$ (but not $\Lambda_2$ or $\Lambda_3$)
if one takes into account $I=1,2,5,6$ ($0<q<\frac{1}{2}, {\vec q}=q\vec{c^*}$, labeling 
of IRREPs as in ref.\cite{Per88}).
For BCCD, which exhibits a soft phonon branch of $\Lambda_3$-symmetry, the primary
contribution is produced by the
SALMs $\{{\altvec W}_{\vec n}^{3\kappa},{\altvec W}_{\vec n}^{7\kappa}\}$.
Additional secondary contributions (transforming according to the IRREPs $\Lambda_3$ and
$\Lambda_2$ which were observed in the fourfold phase\cite{Ezp92}) are accounted for by
a second set of SALMs for each unit (half cell)
($\{{\altvec W}_{\vec n}^{4\kappa}\}$ and $\{{\altvec W}_{\vec n}^{8\kappa}\}$ respectively).
For the
description of the relevant main contribution to structural deformations in BCCD, it 
is sufficient in 
both cases to keep one generalized variable for each set of SALMs, i.e.\ one
has either two ($Q^3_{\vec n},Q^7_{\vec n}$) or four generalized variables 
($Q^3_{\vec n},Q^7_{\vec n}$ {\em and} $Q^4_{\vec n},Q^8_{\vec n}$) per cell.


\subsection{The introduction of the pseudo spin formalism}
The effective potential $\overline \Phi$  is a sum of local half cell potentials 
$\overline{\Phi}^{\text{loc}}$ and harmonic two-variable couplings.
To describe a system with $2^r$ equivalent  minima of 
$\overline{\Phi}^{\text{loc}}$, one
keeps $r$ relevant generalized coordinates $Q^1_{\vec n},\ldots,Q^r_{\vec n}$ per unit.
The positions of the equivalent minima differ only in the signs of
the  $r$ corresponding variables. In general, there are $B$ equivalent
minima per unit at the most.
The ratio of the coupling strength and the height of the local potential barrier 
determines whether the system is more of  the order-disorder
or more of the displacive type.\par

A pseudo spin formulation is derived in the
following way: The interactions are expanded up to second order terms in the variables
\begin{equation}
\label{equation:PSDefinition}
Q^p_{\vec n}= |Q^p_{\vec n}| \sign Q^p_{\vec n} =: |Q^p_{\vec n}| \sigma^p_{\vec n}.
\end{equation}
These are replaced by the averages
$\overline{|Q^p_{\vec n}|} \sigma^p_{\vec n}$
over single wells (taken separately for each $|Q^p_{\vec n}|$). Here 
\begin{displaymath}
\sigma^p_{\vec n}:=\sign Q^p_{\vec n}
\end{displaymath}
is just a shorthand notation for the sign of $Q^p_{\vec n}$. The interaction
between the variables $Q^p_{\vec n}$ and $Q^q_{\vec m}$ becomes
\begin{displaymath}
\overline \Phi^{pq}_{{\vec n}{\vec m}} \cdot
\overline{|Q^p_{\vec n}|} \cdot
\overline{|Q^q_{\vec m}|} \cdot
\sigma^p_{\vec n} \sigma^q_{\vec m}
=: J^{pq}_{{\vec n}{\vec m}} \sigma^p_{\vec n} \sigma^q_{\vec m},
\end{displaymath}
with pseudo spin interactions $J^{pq}_{{\vec n}{\vec m}}= 
\overline \Phi^{pq}_{{\vec n}{\vec m}} \cdot \overline{|Q^p_{\vec n}|} \cdot 
\overline{|Q^q_{\vec m}|}$.
Since they are defined as thermal averages
the latter depend quite naturally on temperature and stresses.
The procedure yields the following pseudo spin Hamiltonian:
\begin{equation}
\label{equation:PSHamiltonian}
{\op H}= \frac{1}{2} \sum_{{\vec n}{\vec m}} \sum_{pq} J^{pq}_{{\vec n}{\vec m}}
\sigma^p_{\vec n} \sigma^q_{\vec m}.
\end{equation}
A switch of $\sigma^p_{\vec n}$ from $+1$ to $-1$ or vice versa is connected with a 
collective motion 
of all particles in the unit according to the corresponding SALM
${\altvec W}^p_{\vec n}$ which is associated with a transition from one well to another.
Because $\sigma$ assumes only the values $+1$ or $-1$, it is called pseudo spin. It
should not be confused with usual spins nor is it localized at
a definite position in the unit. An association of pseudo spins with local normal 
coordinates for the description of soft modes was described in ref.\cite{Ell71}.\par
In this way, the distribution of the variables $Q^p_{\vec n}$  over the wells in the unit is
described by of the operators $\sigma^p_{\vec n}$. The formulation takes 
the actual average displacements and the thermal fluctuations
in the neighborhood of a given equilibrium position into account. A more detailed analysis extending
previous treatments\cite{Gil74,Sta76b} to uniaxially modulated materials and incorporating
both order-disorder and displacive phenomena will be reported in a future publication.\par
It should be mentioned that part of the SALM model
was inspired by an idea presented in a paper\cite{Kra92} describing 
the distortions of four crystallographically
equivalent ammonium tetrahedra in ammonium hydrogen 
oxalate hemihydrate by
four pseudo spins.
This model exhibited the
possibility of incorporating  symmetry considerations into pseudo spin Hamiltonians, 
although the interactions were not properly defined.\par

%
%
\section{Models with continuous variables}
\label{section:contimodels}

Many models proposed so far by various authors for the description of materials with modulated structures
can be connected with the atomistic basis along the lines discussed in section
\ref{section:atomisticmodel}.
In the present section we shall refer to some models with continuous local variables.

\subsection{The Frenkel-Kontorova and DIFFOUR models}
In the Frenkel-Kontorova model\cite{Deh29,Fre38}
a linear chain of atoms with harmonic nearest neighbor couplings in an external periodic
potential incommensurate with the underlying lattice is considered.
The incommensurability of the external potential (period $b$) with the lattice distance $a$
in the potential
\begin{displaymath}
V=\frac{\alpha}{2} \sum_\ell \left[ (x_{\ell+1}-x_{\ell}-a)^2 + \Phi\cdot(1-\cos(2\pi x_\ell/b))\right]
\end{displaymath}
of the Frenkel-Kontorova model produces the frustrations necessary for the appearance of
(in)commensurate structures. In the continuum limit of this model (also known as Frank-van 
der Merwe model\cite{Fra49}) 
the equations of motion lead to the sine-Gordon equation of which the
solutions correspond to a soliton lattice.\par
In the beginning of the 80s, 
Janssen and Tjon proposed a model\cite{Jan81a,Jan82} that they later called
,discrete $\Phi^4$' (DIFFOUR) model following a similar model in field theory.
Ref.\cite{Jan86a} provides an extensive discussion.
The DIFFOUR model is intended as a simple one-dimensional model featuring most of the
typical properties found in modulated crystal phases. 
Its essential ingredient is the competition between short-range interactions.
In its simplest version it 
introduces a linear chain with anharmonic nearest neighbor and harmonic
next nearest and third nearest neighbor interactions.
The frustration mechanism is the
competition of the next  and third nearest neighbor interactions. The reduced potential of the
translation invariant version is given by
\begin{displaymath}
V_{\text{\scriptsize red}}=\sum_\ell\left[ \frac{\alpha}{2}(u_\ell-u_{\ell-1})^2+\frac{\beta}{2}(u_\ell-u_{\ell-2})^2+
\frac{\delta}{2}(u_\ell-u_{\ell-3})^2+\frac{1}{4}(u_\ell-u_{\ell-1})^4 \right],
\end{displaymath}
where $u_\ell$ denotes the displacement of the $\ell$th particle in the chain. $u_\ell$
is scaled such that $V_{\text{\scriptsize red}}$ contains only the necessary model parameters
$\alpha, \beta$ and $\delta$. 
Substituting $x_\ell:= u_\ell-u_{\ell-1}$ yields the
not-translation invariant version:
\begin{displaymath}
{V}_{\text{\scriptsize red}}= \sum_\ell \left[
\frac{A}{2}x_\ell^2+\frac{1}{4}x_\ell^4 + B x_\ell x_{\ell-1} + D x_\ell x_{\ell-2}\right].
\end{displaymath}
This is the potential of an effective Hamiltonian with a local 2-4-potential and nearest and
next nearest neigbor interactions which can be considered as the result of thermal
averaging over the non-relevant variables, while the quantity
$x_\ell$ is interpreted as the relevant degree of freedom for position $\ell$ in the chain.
The coefficients the depend quite naturally on temperature and other control parameters.
The stable structures fulfilling $\partial V_{\text{\scriptsize red}}/\partial x_\ell=0$
and having the smallest free energies 
represent the ground state configurations. A temperature can be introduced either by
identifying the potentials as the respective thermodynamic potentials (e.g.\ free energies) and
assuming
that the parameters of the model are temperature-dependent\cite{Jan82} 
or by starting from a Hamiltonian with the above potential and
treating (a three-dimensional version of) the model
in mean field approximation\cite{Jan83c}.\par
An extension of this model to two degrees of freedom per position\cite{Jan91} leads to
a DIFFOUR model for the description of the phonon spectrum of BCCD\cite{Hli96,Qui96}.
%
\subsection{Chen and Walker's model}
\label{subsection:CWmodel}
Contrary to the approach of the DIFFOUR models starting from a general formulation and
adapting the coefficients to specific substances later,
Chen and Walker use information about the crystal structure and the symmetries
of excitations in uniaxially modulated materials whose high temperature phase 
exhibits the symmetries of the space group $D_{2h}^{16}$ from the beginning. 
Their model was adapted to
the class of \chemical{A_2BX_4} compounds\cite{Che90,Che91b} and to
BCCD\cite{Che91a,Fol91}. The crystal is decomposed in 
equidistant layers with a layer spacing of half the lattice constant along the
direction of modulation. Displacements of the atoms in layer $\ell$ are given by\cite{Che91b}
\begin{displaymath}
v_\ell e_\ell(\Gamma_2) + w_\ell e_\ell(\Gamma_3),
\end{displaymath}
where $e_\ell(\Gamma_2)$ and $e_\ell(\Gamma_3)$ are layer modes belonging to 
the irreducible representations $\Gamma_2$ and $\Gamma_3$ of the two-dimensional
group $C_{2v}$, respectively, from which three-dimensional symmetry modes transforming
like the observed soft mode can be constructed.
Neglecting other layer modes, a Landau-type reduced free energy 
\begin{eqnarray*}
F&=&\sum_\ell\left[\frac{1}{2}a v_\ell^2+\frac{1}{4}v_\ell^4 + \frac{1}{2}a' w_\ell^2 + \frac{1}{4}w_\ell^4 +
b v_\ell^2 w_\ell^2\right]+\\
&&\qquad\frac{1}{2}\sum_\ell\left(J v_\ell v_{\ell+1}+ J' w_\ell w_{\ell+1}\right)+\frac{1}{2}\sum_\ell\left(v_\ell w_{\ell+1}-v_{\ell+1}w_{\ell}\right)
\end{eqnarray*}
invariant under the transformations of the normal phase space group $D_{2h}^{16}$ with
temperature-dependent parameters is introduced. Besides a local energy term, only
nearest neighbor layer interactions are considered. The competition between
the interactions $J$ and $J'$ on the one hand and the antisymmetric interaction term
$v_\ell w_{\ell+1}-v_{\ell+1}w_{\ell}$ on the other hand leads to frustrations and hence
to the occurrence of commensurately and incommensurately modulated structures.\par
Phase diagrams are calculated as planar two-dimensional sections through the
higher-dimensional $(a,a',b,J,J')$-parameter space: the specific structure of
the layer variables $\{v_\ell,w_\ell\}$
minimizing $F$ determines the stable phase at every point of the 
parameter space. The respective space group can be determined from a given profile 
of amplitudes making use of the definition of $e_\ell(\Gamma_2)$ and $e_\ell(\Gamma_3)$.
Up to three different space groups can be found for a given wave number of the modulation.
A path is drawn in the parameter space of the theoretical phase diagram in such a way as to
obtain the same sequence of phases as determined experimentally as a function of
$T$ (for $p=0$) for BCCD\cite{Che91a}.
The expressions for the spontaneous polarization as expansion in terms of the layer variables
can be used to determine the sequence of phases under an external electric field\cite{Fol91}.
Observing that the soft mode in BCCD is an acoustic mode a further layer mode coupled
to the two optic modes and belonging
to the irreducible representation $\Gamma_3$ is introduced\cite{Kap93} in the free energy.
Phonon branches calculated from the latter show a softening of the lowest-lying acoustic mode
for an appropriate choice of the eight reduced parameters.\par
DIFFOUR model and ANNNI model (subsection \ref{subsection:ANNNI}) on the one hand and Chen 
and Walker's model and DIS model (subsection \ref{subsection:DIS})
on the other hand are closely related since in each case the pseudo spin model can be 
considered as
the Ising limit of the model with continuous variables. 

%
\section{Pseudo spin models}
\label{section:PSmodels}
\subsection{The ANNNI model and its extensions}
\label{subsection:ANNNI}
\subsubsection{The model}
The Axial Next Nearest Neighbor Ising (ANNNI) model\cite{Ell61} was
the subject of different excellent review articles\cite{Sel88,Yeo88}.
In the following we will focus on the three-dimensional version and
present the main results as well as some techniques used to examine
the properties of this model. In the ANNNI model only one pseudo spin per
unit $(ijk)$ is considered. The Hamiltonian (without an external field) is given by
\begin{equation}
{\op H} = -J_0 \sum\limits_{ijk} \sigma_{ijk} \left( \sigma_{(i+1)jk} +
\sigma_{i(j+1)k} \right) - J_1 \sum\limits_{ijk} 
\sigma_{ijk} \sigma_{ij(k+1)}
- J_2 \sum\limits_{ijk} \sigma_{ijk} \sigma_{ij(k+2)}.
\label{equation:annniham}
\end{equation}
$J_0 > 0$ is the ferro-type in-plane interaction between nearest
neighbors whereas $J_1$ and $J_2$ are the couplings between nearest
and next nearest neighbors in direction of the modulations respectively. Frustration
effects and therefore modulations arise for $J_2 < 0$ and either sign of $J_1$.
Due to the
invariance of the Hamiltonian (\ref{equation:annniham}) under a
simultaneous transformation $J_1 \rightarrow -J_1$ and 
$\sigma_{ijk} \rightarrow (-1)^k \sigma_{ijk}$, 
it is only necessary
to examine the case $J_1 > 0$. Using reduced units, the ground state
is given by the ferro phase for $\kappa < \frac{1}{2}$ and
the $\left< 2 \right>$-phase for $\kappa > \frac{1}{2}$ with 
$\kappa = - \frac{J_2}{J_1}$. For the $\left< 2 \right>$-phase the
spin structure repeats itself after four layers and is given by
$++--$, i.e.\ two layers where all the pseudo spins have the value +1
followed by two layers with $-1$ pseudo spins. The two ground phases are 
separated by a multiphase point (MP) where an infinity of phases
is degenerate.\par
The A3NNI (Axial Third Nearest Neighbor Ising) model\cite{Sel85,Bar85,Ran85}
differs from the usual ANNNI model by considering a third neighbor
interaction in {\bf c}-direction. This extension leads to a 
two-dimensional ground state phase diagram where some of the phases
are separated by multiphase lines (ML). The two phases $\left< 12
\right>$ and $\left< 3 \right>$ occupy now a finite stability region
at $T = 0$.\\
Starting point for the ELII (Effectively Long-range Interaction Ising)
model\cite{Vil80,Mas83} is the coupling between two degrees of freedom
(one degree of freedom is considered to be an Ising spin) which
leads to indirect effective long-range oscillating interactions
$J_r = \sum\limits_q J_q e^{{\mathrm i}qr}$ between the pseudo spins, $r$ being the
number of layers separating the interacting pseudo spins.

\subsubsection{The mean field approximation}
In the mean field approximation (MFA)\cite{Bak80,Sel84b}
it is assumed that every spin interacts
only with a mean field produced by the thermal averages of all spins 
interacting with it. This is equivalent to neglecting spin fluctuations
or approximating the density operator by a product of single spin density
operators. The free energy 
of the ANNNI model derived in this approximation (apart from constant factors)
\begin{eqnarray}
F & = & \sum\limits_k \left[ - 2 J_0 S_k^2 - J_1 S_k S_{k+1} -J_2 S_k S_{k+2}
\right. \nonumber \\
& & \hspace*{0.8cm} \left. + k_BT \left( \left(1 + S_k \right) \ln \left(1 + S_k \right) +
\left(1 - S_k \right) \ln \left(1- S_k \right) \right) \right]
\nonumber 
\end{eqnarray}
is a function of the the mean field value 
$S_k = \left< \sigma_{ijk} \right>$ of the $k$-th layer. Minimizing the
free energy leads to the equilibrium equations
\begin{equation}
S_k = \tanh \left[ 4 J_0 S_k + J_1 \left( S_{k-1} + S_{k+1} \right)
+ J_2 \left( S_{k-2} + S_{k+2} \right) \right].
\label{equation:equilibrium}
\end{equation}
The critical line 
\begin{displaymath}
k_B T_{\text{crit}} ( \kappa ) = \left\{ \begin{array}{ll}
4J_0+ 2(1 - \kappa) J_1  & \qquad\kappa \le \frac{1}{4} \\[3mm]
4J_0+ (2 \kappa + \frac{1}{4 \kappa} ) J_1 & \qquad\kappa \ge \frac{1}{4}
\end {array} \right.
\end{displaymath}
which separates the disordered para phase from the ordered phases and the critical
wave vector
\begin{equation}
\label{equation:qcrit}
q_{\text{crit}} = \left\{ \begin{array}{ll}
0 &\qquad \kappa \le \frac{1}{4} \\[3mm]
\arccos \frac{1}{4 \kappa} &\qquad \kappa \ge \frac{1}{4}
\end {array} \right.
\end{equation}
are obtained by linearizing the equilibrium equations 
(\ref{equation:equilibrium}). At the Lifshitz point ($\kappa = \frac{1}{4}$)
the para phase, the ferro phase and the modulated phases meet.
Figure (\ref{figure:annnimeanfield}) (from ref.\cite{Sel84b}) shows a typical mean field phase
diagram with the main commensurate phases derived for the case $J_0 = J_1$. 
Mean field phase diagrams of the A3NNI\cite{Ran85} and of the ELII model\cite{Mas83} have also
been published.\\
Duxbury and Selke\cite{Dux83,Sel84b} verified for
the ANNNI model the existence of structure
combination branching processes responsible for the creation of new phases
at finite temperatures. In this process two neighboring stable phases $\left< A
\right>$ and $\left< B \right>$ produce a new phase $\left< A B \right>$
at a definite branching point. At higher temperatures this process repeats
itself leading for example to the appearance of the phases
$\left< A^n B \right>$ at different branching points. 
The order of the phase transition changes at the accumulation points
of the branching processes
from first to second order.
The coordinates of the accumulation point of the $\left< 2 3^n \right>$
phases were determined by numerical extrapolation\cite{Sel84b}. Using a
fixed point expansion the positions of accumulation points were 
calculated numerically and analytically
for very different series of phases\cite{Sie89a,Ten90a,Ten92,Jen94a,Jen96}. This
fixed point expansion will be the subject of the next section.\\
MFA seemed to indicate the existence of partially
disordered\cite{Yok81,Nak89} and asymmetric\cite{Yok91} phases
for weak in-layer coupling $J_0$ (i.e.\ $J_0 \le 0.3 J_1$).
In order to further examine this behavior Nakanishi\cite{Nak92}
used the mean field transfer matrix method (MFTM) where the interactions
in {\bf c}-direction are treated exactly, while the interactions in the layers are
approximated by mean fields. The problem is reduced to 
one dimension with the free energy
\begin{displaymath}
F = \sum\limits_k \left[ - \frac{1}{2} J_0 S_k^2 + h_k S_k \right]
- k_B T \ln Z
\end{displaymath}
where the partition function $Z= \mbox{Tr} ~ \exp \left[ 
\beta \sum\limits_k \left(
J_1 \sigma_k \sigma_{k+1} + J_2 \sigma_k \sigma_{k+2} + h_k \sigma_k \right)
\right]$ can be calculated by use of transfer matrices. $h_k$ is the
mean field of the $k$-th layer. Minimizing the free energy with respect 
to the calculated thermal average $S_k$ leads to a system of nonlinear
equations which have to be solved selfconsistently. As a result
Nakanishi showed that the free energies of the partially disordered
and the asymmetric states are always higher than the free energy of the
symmetric state and concluded that the former states were an artifact
of the MFA. Monte Carlo calculations\cite{Rot93} supported this claim.

\subsubsection{The fixed point expansion}
\label{subsubsection:fixedpointexpansion}
The fixed point expansion method
was developed for the calculation of profiles, free energies and
interaction energies of discommensurations in the frame of the ANNNI
model\cite{Axe81,Jan83a,Hog84,Ten88b,Sie89a,Sie89b,Ten90a}. 
The mean field equilibrium equations (\ref{equation:equilibrium}) 
are reformulated as a four-dimensional mapping ${\altvec P}^{k+1} 
= {\altvec h} ( {\altvec P}^k )$ for the vectors ${\altvec P}^k = ( S_{k-2},
S_{k-1},S_k,S_{k+1} )$\cite{Sie89a}. The pseudo spin 
profile ${\altvec P}^{\gamma k}$ of 
a phase $\gamma$ with the periodicity $N$ verifies the relation
${\altvec P}^{\gamma k} = {\altvec P}^{\gamma \, k + N} = {\altvec h}^N 
( {\altvec P}^{\gamma k} )$ and corresponds, thus, to an iterated fixed point
of the mapping.\par
Since a discommensuration is a localized defect the changes in the
pseudo spin profile are the largest in the immediate vicinity of this defect.
This region is treated numerically by means of the original (not linearized)
equilibrium equations. The large number of pseudo spins whose numerical 
treatment would be cumbersome is conveniently handled by an expansion
in the asymptotic region, however, about the fixed point leading to the linearized mapping
${\altvec \Delta}^{k+1} = {\altmat H}^k {\altvec \Delta}^k$
for the deviations ${\altvec \Delta}^k = {\altvec P}^{\gamma k} -
{\altvec P}^k$. The matrices ${\altmat H}^k$ are symplectic matrices, whose special
properties allow a largely analytic derivation of the profiles, the free energies and
the interactions of such defects. Of special
importance is the (symplectic) product ${\altmat G}^k = \prod_{\mu = 1}^n
{\altmat H}^{k+n-\mu}$ where the period $n$ of the matrices ${\altmat H}^k$ 
(i.e.\ ${\altmat H}^{k+n} = {\altmat H}^k$) is $n = N$ or $n = \frac{N}{2}$ 
depending on the symmetry of the profile ${\altvec P}^{\gamma k}$.\\
The asymptotic behavior of the deviations is determined in terms of the
product matrix ${\altmat G}^k$ in the form ${\altvec \Delta}^{k+n} = {\altmat G}^k \,
{\altvec \Delta}^k$ and therefore by the eigenvalues 
$\lambda_{\nu}$ of ${\altmat G}^k$. Figure (\ref{figure:fixpoint})
shows the different fixed point types resulting from the positions of
the four eigenvalues in the complex plane. The approach to the
fixed point is thus monotonic for fixed point type 2 ($2'$) and
oscillatory for type 3 (see ref.\cite{Sie89a} for a detailed discussion 
of the fixed point types).\par
Analytical results for the free energy $\Sigma$ of one discommensuration
or the interaction energy $W_2 (d)$
of two discommensurations (which depends on the distance $d$ 
between the two discommensurations) can be derived. If the fixed point
type is 2 ($2'$) the interaction $W_2 (d)$ between two discommensurations
is a monotonic function of $d$. For the fixed point type 3 the
interaction is oscillatory\cite{Sie89a}.\par
The fixed point expansion is very useful for the determination of the order of the 
transitions and for a
(numerical and analytical) calculation of the positions of accumulation
points\cite{Ten90a,Ten92,Jen94a,Jen96}. Suppose two neighboring stable commensurate
phases $\gamma$ and $\gamma'$ are separated by a transition line.
The phase $\gamma'$ can be viewed as resulting from a periodic
arrangement of $Z$ adequate discommensurations in the phase $\gamma$ separated by
$d_1, \ldots , d_Z$ ($d_{\nu + Z} = d_{\nu}$) layers. The difference
of the free energies per layer is
\begin{displaymath}
F = F^{\gamma} - F^{\gamma'} = Z \, \Sigma + W(d_1, \ldots , d_Z) \simeq
Z \, \Sigma + \sum\limits_{\nu=1}^Z W_2 ( d_{\nu})
\end{displaymath}
where we use the fact that usually, due to the rapid exponential decay of 
$W_2 ( d )$, only nearest neighbor discommensuration interactions are to
be considered. The phase transition between the phases $\gamma$ and
$\gamma'$ takes place when $F$ is equal to zero. If, at the transition,
the interaction is monotonic [fixed point type 2 ($2'$)]
the transition is of second order and
occurs where $\Sigma$ is zero, the distance $d$ of neighboring
spontaneously formed discommensurations being infinite. If, on the
contrary, the interaction is oscillatory (fixed point type 3)
the transition is of first order
and takes place for  positive $\Sigma$. The distance of neighboring
discommensurations is then finite.\\
For a phase $\gamma$ three different lines meet at the accumulation
point: the phase boundary, the line $\Sigma = 0$ and the line
separating the fixed point types 3 and 2 ($2'$). This enabled 
Tentrup, Jenal, and Siems to find numerically the accumulation points
of the series $\left< 2^k 3 \right>$\cite{Ten92}, $\left< 2 3^k \right>$\cite{Jen94a} 
and $\left< 2^{k+1} 3 2^k 3 \right>$\cite{Jen96}. A low
temperature mean field analysis allowed to derive analytical
approximations\cite{Ten92,Jen96} which render the coordinates of
the accumulation points well.

\subsubsection{The low temperature series expansion}
In their seminal work Fisher and Selke \cite{Fis80,Fis81} derived
the low temperature behavior of the ANNNI model by
analyzing an exact series expansion of the free energy. Herefore,
they used the concept of structure variables $l_{\mu}$,
thus characterizing each possible ground state by
an unique set of these variables. The variable $l_{\mu}$ is defined by
$l_{\mu} = \frac{L_{\mu}}{L}$ where $L_\mu$ is the number of
band sequences of type $\mu$ in a given structure in a lattice of $L$
layers \cite{Fis81}. The free energy can then be written in the form
\begin{displaymath}
F = \sum\limits_{\mu} a_\mu l_{\mu}
\end{displaymath}
where the sum is over all possible band squences. The parameters
$a_\mu$ are expanded in the form 
\begin{displaymath}
a_\mu = \sum\limits_k a_\mu^{(k)},
\end{displaymath}
the contributions $a_\mu^{(k)}$ resulting from the $k$th order term
in the series expansion of the free energy. 
In the $k$th order term of the free energy enter all terms where
the values of $k$ spins differ from their ground state value.\\
In their analysis of the different orders of the free energy
Fisher and Selke concluded\cite{Fis80,Fis81} that the infinity of phases
with the phase symbol $\left< 2^k 3 \right>$ are stable at low 
temperatures. Later Fisher and Szpilka refined this expansion using
a transfer-matrix method\cite{Fis87b}. They showed that among the periodic
phases $\left< 2^k 3 \right>$ only those with $1 \le k \le k_{\text{max}}$ appear
at any fixed small temperature. Since $k_{\text{max}}$ diverges as $T \longrightarrow
0$ all phases of the form $\left< 2^k 3 \right>$ indeed spring from
the MP. Similarly they concluded that the mixed phases 
$\left< 2^k 3 2^{k+1} 3 \right>$  appear for $k^{(1)}(T) \le k \le
k^{(2)}(T)$. Since $\frac{k^{(1)}}{k_{\text{max}}} \longrightarrow 0$ and
$\frac{k^{(2)}}{k_{\text{max}}} \longrightarrow 1$ in the limit 
$T \longrightarrow 0$ phases $\left< 2^k 3 2^{k+1} 3 \right>$ appear
arbitrarily close to the MP.\\
This low temperature series expansion technique has also been applied
to other models exhibiting modulations, for example to the A3NNI
model\cite{Bar85}, the three-state chiral clock model\cite{Yeo84}, the generalized
$p$-state chiral clock model\cite{Yeo82}, a six-state clock model with
next nearest neighbor interactions\cite{Sen93}, or the DIS (Double
Ising Spin) model\cite{Ple96}. This latter model will be the subject
of subsection \ref{subsection:DIS}.

\subsubsection{The high temperature series expansion}
In the high temperature series expansion\cite{Red77a,Red77b,Oit85}
the wavevector dependent susceptibility 
\begin{displaymath}
\chi ( q ) = \sum\limits_j \left< \sigma_0 \sigma_j \right> e^{2 \pi 
{\mathrm i} q z_j}
\end{displaymath}
is considered,
where the sum is over all correlations and $z_j$ is the number of lattice
spacings in $\bf c$-direction. Writing the correlation function
\begin{displaymath}
\left< \sigma_0 \sigma_j \right> = \frac{\mbox{Tr} \left( e^{- \beta H}
\sigma_0 \sigma_j \right)}{\mbox{Tr} \left( e^{- \beta H} \right)}
\end{displaymath}
as an high temperature expansion one can construct a $\chi$-series
for any wavenumber. As a result the Lifshitz point (for the case
$J_0 = J_1$) was found to occur for $\kappa = 0.27$\cite{Red77b,Oit85}.
The critical wavenumber and the critical temperature as function 
of $\kappa$ were also calculated. In ref.\cite{Mo91}  the critical exponents
for the different regions along the para phase boundary (i.e., for the
para phase-ferro phase transition, the para phase-modulated region 
transition, as well as for the multicritical Lifshitz point) were 
calculated for the ANNNI model formulated on a sc or fcc lattice using
a series expansion. It was found, in agreement with the results
of renormalization group theory\cite{Gar76,Dro76}, that the critical
exponents of the phase transition from the para phase to the
modulated region are given by the exponents of the 3d-XY model,
whereas the critical behavior of the transition from the para to
the ferro phase is that of the 3d-Ising model.

\subsection{The AANNDI model}
\label{subsection:AANNDI}
The Axial Antisymmetric Nearest Neighbor Double Ising (AANNDI) 
model\cite{Kur94} has originally been formulated\cite{Kur86} as an
order-disorder mechanism of phase transitions for the compounds of
the \chemical{A'A''BX_4}-family. Two pseudo spins were introduced,
describing the orientational states of a \chemical{BX_4} tetrahedron. 
The four discrete orientational states are labeled by 
$\sigma^A = \pm 1$ (apex up or down) and $\tau^A = \pm 1$
(turn to the right or left). Remarking that the pseudospins $\tau^A$,
$\sigma^A$ and the product $\tau^A \sigma^A$ transform according to 
three different onedimensional representations of the local symmetry
group, the general Hamiltonian 
\begin{eqnarray}
{\op H} & = & J' \sum\limits_{\text{nn} ~ \text{in}} \sigma^A_i \sigma^A_j + K' \sum\limits_{\text{nn} ~ \text{in}}
\tau^A_i \tau^A_j + L' \sum\limits_{\text{nn} ~ \text{in}} \sigma^A_i \tau^A_i \sigma^A_j \tau^A_j
+ J \sum\limits_{\text{nn} ~ \text{out}} \sigma^A_i \sigma^A_j \nonumber \\
& & + K \sum\limits_{\text{nn} ~ \text{out}}
\tau^A_i \tau^A_j \nonumber
+ L \sum\limits_{\text{nn} ~ \text{out}} \sigma^A_i \tau^A_i \sigma^A_j \tau^A_j
+ \frac{1}{2} M' \sum\limits_{\text{nn} ~ \text{in}} \sigma^A_i \sigma^A_j \left( \tau^A_i
- \tau^A_j \right) \nonumber \\
& & + \frac{1}{2} M \sum\limits_{\text{nn} ~ \text{out}} \tau^A_i \tau^A_j
\left( \sigma_i - \sigma_j \right)
\label{equation:generalaanndi}
\end{eqnarray}
was derived by reta\text{in}\text{in}g only symmetry-allowed nearest neighbor
coupl\text{in}gs between $\tau^A$, $\sigma^A$ and $\tau^A \sigma^A$. The sums
$\text{nn} ~\text{in}$ and $\text{nn} ~\text{out}$ are over nearest neighbors \text{in} the planes and
perpendicular to the planes respectively.\\
The AANNDI model is derived from eqn.\  (\ref{equation:generalaanndi})
by retaining the terms
\begin{equation}
{\op H} = J \sum\limits_l \sigma^A_l \sigma^A_{l+1} + K \sum\limits_l \tau^A_l
\tau^A_{l+1} + \frac{1}{2} M \sum\limits_l \tau^A_l \tau^A_{l+1} \left(
\sigma^A_l - \sigma^A_{l+1} \right)
\label{equation:aanndi}
\end{equation}
where $l$ labels the different layers. In-plane couplings assuring
ferro or antiferro orderings in the planes are assumed but not taken
explicitly into account. The Hamiltonian (\ref{equation:aanndi}) is
therefore only a onedimensional one which is examined in 
mean-field approximation\cite{Kur90}. In this analysis the product\cite{Kur92b}
$\omega^A_l = \tau^A_l \sigma^A_l$ was taken to be independent of 
$\tau^A_l$ and $\sigma^A_l$ thus leading to a mean-field free energy
depending on the three layer order parameters $\left< \tau^A_l \right>$,
$\left< \sigma^A_l \right>$ and $\left< \omega^A_l \right>$. A closer
look reveals that in the equilibrium equations $\left< \sigma^A_l \right>$
decouples from the other order parameters. Therefore the term proportional to
$J$ can be neglected\cite{Kur94}.\\
It should be noticed that, retaining the antisymmetrical interactions in
the planes (cf.\ eqn.\ (\ref{equation:generalaanndi})), one obtains the PANNDI
(Planar Antisymmetrical Nearest Neighbor Ising) model suitable for
the investigation of twodimensional modulations in
\chemical{A'A''BX_4}-compounds.

\subsection{The Double Ising Spin (DIS) model}
\label{subsection:DIS}
The Double Ising Spin or DIS model has been proposed by us\cite{Ple94,Neu94a,Ple96} 
as a simple model for the investigation of uniaxially modulated
materials with the para phase symmetry group $D^{16}_{2h}$. Its Hamiltonian can be
derived from the general Hamiltonian (\ref{equation:generalaanndi})
by introducing the two spins $\tau \equiv \tau^A$ and $\sigma \equiv
\sigma^A \tau^A$. Retaining only symmetric in-layer couplings
(as we are not interested in twodimensional modulations) we finally
obtain the DIS Hamiltonian
\begin{eqnarray}
{\op H} &=& K \sum\limits_{ijk} \tau_{ijk} \tau_{ij(k+1)} + L \sum\limits_{ijk}
\sigma_{ijk} \sigma_{ij(k+1)} + \frac{M}{2} \sum\limits_{ijk} \left(
\sigma_{ijk} \tau_{ij(k+1)} - \tau_{ijk} \sigma_{ij(k+1)} \right) \nonumber \\
& & + J \sum\limits_{ijk} \tau_{ijk} \left( \tau_{(i+1)jk} + \tau_{i(j+1)k}
\right) + J' \sum\limits_{ijk} \sigma_{ijk} \left( \sigma_{(i+1)jk} +
\sigma_{i(j+1)k} \right) 
\label{equation:dis}
\end{eqnarray}
Comparing it to the AANNDI Hamiltonian (\ref{equation:aanndi}) reveals 
that the unimportant term $J$ has been dropped whereas the term $L$ is
considered. Furthermore in-layer couplings are explicitly taken into
consideration giving way to a three-dimensional model. In the
mean-field treatment we have only two order parameters per layer
($\left< \tau \right>$ and $\left< \sigma \right>$) compared to 
the three order parameters of the AANNDI model.\\
Frustration and therefore modulations arise 
in {\bf c}-direction because of the antagonistic
effects of the symmetric nearest neighbor interactions $K$ and $L$
on the one hand and the antisymmetric interaction $M$ on the other hand.
The in-layer couplings $J$ and $J'$ are ferro couplings.\\
In the ground state phase diagram (figure \ref{figure:disground})
five different phases with a twodimensional stability region are
separated by multiphase lines at which infinitely many different phases
are degenerate. In the following we will only consider
the multiphase lines $K-L = -M$
(line 1) and $K+L=-M$ (line 2). The lines 3 and 4 are obtained from the
lines 1 and 2 by the symmetry of the Hamiltonian.\\
The global temperature-interaction phase diagram is 
examined in the mean-field approximation. One obtains very different
twodimensional sections of the phase diagram depending on whether, at $T = 0$, 
the line 1 (figure \ref{figure:diagram1}) or the line 2 
(figure \ref{figure:diagram2}) is intersected. In the first case the
modulated phases are only encountered at rather high temperatures, in
the second case {\em all} examined phases reach to low temperatures
and seem to converge to the multiphase point. Only some of the commensurate
phases are shown, the high-commensurate and the incommensurate phases
filling up the shaded areas.\\
Using the fixed point expansion (cf.\ subsection \ref{subsubsection:fixedpointexpansion})
we examined the
processes leading to the appearance of new phases\cite{Ple94}. In the case 1 
the existence of structure combination branching processes and of
accumulation points of these branchings has been verified, i.e.\ new phases
are created by the same processes as in the ANNNI model. In the case 2
no branching points nor accumulation points have been found. The phase 
creation process is therefore different from the process in the 
ANNNI model.\\
In order to further clarify this different behavior we performed a
low temperature series expansion of the free energy\cite{Ple96} following the
ideas by Fisher and Selke\cite{Fis80,Fis81} for the investigation of the ANNNI
model (i.e.\ of a model with one-component spins). This expansion reveals that
in case 1 only four of the infinity of phases degenerate at the
MP are stable at low temperatures. All other modulated phases are created
by branching processes at higher temperatures. For case 2 it follows from
the low temperature expansion that an infinity of different phases spring
from the MP. At higher (but still small) temperatures some of the 
high-commensurate phases disappear. A close inspection of the matrices
and vectors entering the computation of the free energies\cite{Ple97b}
reveals that in the limit $T \longrightarrow 0$ the DIS model exhibits
a complete devil's staircase for case 2.\par 
The critical properties of the model can be derived
by writing the Hamiltonian (\ref{equation:dis}) in matrix form.
A special local transformation then leads to a Hamiltonian where the
interaction terms are given by scalar products between 
two-component vectors\cite{Ple97b}. For the case $K<0$ and $L<0$ the partition function
of this
Hamiltonian can then be transformed to the partition function of an
effective Ginzburg-Landau-Wilson-Hamiltonian for which one can show
that it has the critical behavior of the 3d-XY model.\par
A connection with experimental results obtained for actual crystals is established by
applying the method described in section \ref{section:application} to the DIS model.
For BCCD, for example, this
leads to a description\cite{Neu94a} which is at least  as good as
the description by the ANNNI model, the latter being a special case of
the former\cite{Ple97a}. 

\subsection{Interrelations of various pseudo spin models}
A general $r$-spin model containing $r$ spins per cell 
$\sigma^1_{\vec n}, \ldots, \sigma^r_{\vec n}$ can also be
described as a $q = 2^r$-state model.  The state variable $p_{\vec n}  =
1, \ldots , 2^r$ and the spin variables $\underline{\sigma}_{\vec n} =
\left( \sigma^1_{\vec n}, \ldots, \sigma^r_{\vec n} \right)$ are,
for example, connected by the equation\cite{Ple97a}
\begin{displaymath}
p_{\vec n}  = 1 + \frac{1}{2} \sum\limits_{i=1}^r \left( \sigma^i_{\vec n}
+ 1 \right) \, 2^{i-1}.
\end{displaymath}
Special cases of $r$-spin models are, e.g., $2^r$-state Potts models.\par
Pseudo spin systems in the narrower sense are described by
Hamiltonians of the form
\begin{displaymath}
{\op H} = \sum\limits_{{\vec n},{\vec m}} \Lambda^{ij}_{{\vec n}{\vec m}}
\sigma^i_{\vec n} \sigma^j_{\vec n}.
\end{displaymath}
For uniaxially modulated systems with nearest neighbor interactions
in the planes perpendicular to the direction of the modulations
this Hamiltonian may be rewritten as
\begin{displaymath}
{\op H} = {\sum_{\left< {\vec n},{\vec m} \right>}}^{\!\!\!\!\perp}
V_{\vec nm}^{ij} \sigma^i_{\vec n} \sigma^j_{\vec m} +
{\sum_{{\vec n},{\vec m}}}^{\|} W_{\vec nm}^{ij}
\sigma^i_{\vec n} \sigma^j_{\vec m}.
\end{displaymath}
The matrix $\altmat V$ denotes couplings perpendicular to
the modulation direction, the sum running over all nearest neighbor
pairs $\left< {\vec n},{\vec m} \right>$ in the planes, whereas
$\altmat W$ denotes couplings along the direction of the 
modulations.\par
In the previous sections we discussed three different pseudo spin
models: the ANNNI model (section \ref{subsection:ANNNI}), 
the AANNDI model (section \ref{subsection:AANNDI}))
and the DIS model (section \ref{subsection:DIS}). Whereas the ANNNI model is an
one-spin model, the DIS and the AANNDI model are, in their original
formulation, two-spin models. However, in the treatment of the AANNDI model
Kurzy{\'n}ski considered three independent order parameters
$\left< \tau^A \right>$, $\left< \sigma^A \right>$ and $\left< \omega^A
\right> = \left<
\tau^A \sigma^A \right>$, i.e.\ the AANNDI model is treated as a model with  the
three pseudo spins 
$\tau^A$, $\sigma^A$ and $\omega^A$. The DIS model is, on the contrary,
treated as a model with two pseudo spins $\tau$ and $\sigma$ per cell.\par
Both the AANNDI and the DIS model can be derived from the general
Hamiltonian (\ref{equation:generalaanndi}).\par
The DIS model represents a model which is somewhat more general than the ANNNI model
but nevertheless so simple, that it allows the derivation of explicit results
(dependent on a few well defined local interactions) which can be compared to
experimental data. This powerful model contains some other much discussed models
as special cases:\par
Since a general $r$-Ising spin model (i.e.\ a model, where $r$ spins describe
the state of every unit) can also be considered to be a $2^r$-state
model\cite{Ple97a}
(i.e.\ a variable $p = 1, \ldots , 2^r$ describes the state of every 
unit), it follows that the DIS model can be formulated as a four-state
model. For the special case $K = L < 0$ and $J = J' < 0$ the
DIS Hamiltonian can be rewritten in the form\cite{Ple97a}
\begin{displaymath}
{\op H} = - \zeta_0 {\sum\limits_{<{\bf n},{\bf m}>}}^{\!\!\!\!\perp} ~ \cos \left(
\frac{\pi}{2}
\left( p_{\bf n} - p_{\bf m} \right) \right) - \zeta {\sum\limits_{<{\bf n},{\bf
 m}>}}^{\!\!\!\!\parallel}
\cos \left( \frac{\pi}{2} \left( p_{\bf n} - p_{\bf m} + \Delta \right) \right),
\end{displaymath}
where we introduced the variables 
\begin{displaymath}
\zeta_0 := - 2 J~~~,~~~ \zeta := 
\sqrt{M^2 + 4 K^2} ~~\mbox{and}~~ \Delta := - \frac{2}{\pi} \arctan 
\left( \frac{M}{2 K} \right).
\end{displaymath}
This is exactly the Hamiltonian of the four-state chiral
clock ($CC_4$) model\cite{Yeo82}. It follows from the exact expansion
of the free energy for the special case $K = L < 0$ of the DIS model
that in the limit $T \longrightarrow 0$ the $CC_4$ model exhibits
a complete devil's staircase. This is in marked contrast to an earlier
statement\cite{Yeo82} that only specific phases
spring from the MP. A detailed analysis of this rather complicated
calculation reveals\cite{Ple97b} that in the earlier investigation the treatment of the
in-layer interactions was erroneous, thus leading to wrong expressions
for the free energies. In fact, recent Monte Carlo simulations\cite{Sch96a}
of the $CC_4$ model show the existence of modulated phases not 
predicted by ref.\cite{Yeo82} but in accordance with our results.\par
It can further be shown\cite{Ple97a} that a special case of the
DIS model (two pseudo spins per cell with nearest neighbor interactions)
can be mapped exactly onto the ANNNI model (one pseudo spin per cell 
with nearest and next nearest neighbor interactions). At the first sight this is
surprising, since with respect to the range of direct interactions the ANNNI model
is more general. For the special case considered  no
direct couplings between neighboring $\sigma$-spins are retained, 
i.e.\  $L = 0$ and $J' = 0$ in eqn.\  (\ref{equation:dis}). The $\sigma$-spins
then mediate an indirect next nearest neighbor coupling between the
$\tau$-spins. The resulting relation between the coupling parameters of the
two models is temperature-dependent. Its explicit analytic form is\cite{Ple97a} 
\begin{eqnarray*}
J_0 & = & - J, \\
J_1 & = & - K, \\
J_2 & = & - \frac{1}{2 \beta} \ln \cosh \left( \beta M \right) .
\end{eqnarray*}

\section{Connection of model calculations with experimentally determined 
properties of actual materials}
\label{section:application}

\subsection{Theoretical derivation of phase diagrams in terms of experimental
control parameters}
\label{subsection:theoderivedpd}
All the reviewed theoretical models for the description of
uniaxially modulated materials have in common that the displayed
phase diagrams are twodimensional sections of a higher dimensional
space spanned by the model parameters (for example reduced temperature,
reduced interactions etc.). The experimental phase diagrams, however,
are spanned by external quantities like temperature, pressure
or applied fields. It is thus necessary to find a mapping between
the experimental quantities and the theoretical parameters in order
to compare the experimental and theoretical phase diagrams. In
the following we will present a procedure for transforming
a two-dimensional theoretical phase diagram from model parameters to
temperature-pressure variables. This mapping has originally been
formulated for the ANNNI and the DIS model\cite{Ten90b,Neu94b}
in order to describe BCCD. Since the experimental wave numbers $q$ are
given as multiples of $2\pi/c$, the assignment of units to half
cells yields (for the ANNNI model) $q=\nu/(Z_1+Z_2+\ldots+Z_\nu)$ for the relation between the phase symbol
$\langle Z_1 Z_2\ldots Z_\nu\rangle$ and $q$. We will now present a more general
approach transforming arbitrary theoretical phase diagrams into
temperature-pressure phase diagrams.\par
The method requires the knowledge of the thermal expansion coefficients, the elastic constants,
and information on the critical line $T_{\text{crit}}(p_{\text{crit}})$ 
separating the normal phase from the lower-symmetry phases in the $T$-$p$ diagram. 
From the determination of the critical wavenumber
for different pressures\cite{Ao89} a fit $q_{\text{crit}}(p_{\text{crit}})$ can be found.\par
Let the two quantities spanning the
considered two-dimensional theoretical diagram be $\kappa$ and $\theta$.
In the ANNNI model, for example, $\kappa$ is given by $\kappa =
- \frac{J_2}{J_1}$ whereas $\theta$ is the reduced temperature
$\theta = \frac{k_BT}{J_1}$. The critical line $\theta_{\text{crit}}(\kappa)$ 
and the critical wavenumber $q_{\text{crit}}(\kappa)$ result
from an analytical calculation (ANNNI, AANNDI, DIS model) or from a
numerical fit (DIFFOUR, Chen and Walker's model).\par
In the first step the critical line $\theta_{\text{crit}}(\kappa_{\text{crit}})$ is transformed
into a critical $T$-$p$-line. For a given $\kappa = \kappa_0$
the critical values $q_{\text{crit}}(\kappa_0)$ and $\theta_{\text{crit}}(\kappa_0)$ are obtained.
On the other hand, as the wavenumber $q_{\text{crit}}$ is now known, the
values $p_{\text{crit}}(q_{\text{crit}})$ and $T_{\text{crit}}(p_{\text{crit}}(q_{\text{crit}}))$ for the critical pressure can be
derived, thus leading to a theoretical
$T$-$p$ critical line.\par
In the next step we consider an arbitrary point $\left( \kappa, \theta \right)$.
The temperature coordinate is obtained by the simple transformation rule
$k_B T = \frac{k_B T_{\text{crit}}}{\theta_{\text{crit}}} \theta$. Depending on the models
this rule may be replaced by a more sophisticated approach. For simplicity, 
we assume that the model interactions depend only via
the mean lattice constant on temperature and pressure, 
i.e.\ $J_k = J_k \left( \overline{a}(T,p) \right)$\cite{Ten90b,Neu94b}.
Under this assumption the interactions are constant along
lines where $\overline{a}$ is constant. Lines with constant 
$\overline{a}$ are especially lines with constant volume. Expanding the
volume about its value for a given temperature $T_0$ and pressure $p_0$ leads
to the equation
\begin{eqnarray}
V(T,p) & = & V(T_0,p_0) + \frac{\partial V}{\partial T} (T_0,p_0) \cdot
\left( T - T_0 \right) + \frac{\partial V}{\partial p} (T_0,p_0) \cdot
\left( p - p_0 \right) \nonumber \\
& & + \frac{1}{2} \frac{\partial^2 V}{\partial T^2} (T_0,p_0) \cdot
\left( T - T_0 \right)^2 + \frac{1}{2} \frac{\partial^2 V}{\partial p^2}
(T_0,p_0) \cdot \left( p - p_0 \right)^2 \nonumber \\
& & + \frac{\partial^2 V}{\partial T \partial p} (T_0,p_0) \cdot
\left( T - T_0 \right) \left( p - p_0 \right) \nonumber
\end{eqnarray}
or
\begin{eqnarray}
v & = & \frac{V(T,p) - V(T_0,p_0)}{V(T_0,p_0)} \nonumber \\
& = & 3 \overline{\alpha} \cdot
\left( T - T_0 \right) - k \cdot \left( p - p_0 \right) +
 \frac{3}{2} \frac{\partial \overline{\alpha}}{\partial T} \cdot
\left( T - T_0 \right)^2 \nonumber \\
& & - \frac{1}{2} \frac{\partial k}{\partial p} \cdot 
\left( p - p_0 \right)^2 - \frac{\partial k}{\partial T} \cdot
\left( T - T_0 \right) \left( p - p_0 \right) 
\label{equation:transform}
\end{eqnarray}
where we introduced the thermal expansion coefficient $\overline{\alpha}
= \frac{1}{V} \frac{\partial V}{\partial T}$ 
and the compressibility $k= - \frac{1}{V} \frac{\partial V}{\partial p}$.
The value of $v$ is determined by the intersection point of
the line of constant volume with the para phase boundary, 
i.e.\ $v = \frac{V(T_{\text{crit}},p_{\text{crit}}) - V(T_0,p_0)}{V(T_0,p_0)}$.
Inserting the temperature $T$ in eqn.\ (\ref{equation:transform})
finally leads to the respective pressure.\par
Figure \ref{figure:ANNNIpTPD} shows the theoretical 
temperature-pressure phase diagram derived from the ANNNI model
for the description of BCCD\cite{Neu94a}. This diagram shows the same topology as 
the experimental diagram, similar extensions of the modulated phases,
and the observed branchings.  
The phase diagram for the DIS model is transformed in a similar way as the 
ANNNI-phase diagram\cite{Neu94a}.\par
The observed\cite{Kir93} increase of transition temperatures with
uniaxial compressional strains $-\epsilon_{ii}$ corroborates the
assumption that the interactions depend on $p$ and $T$ only via the 
lattice constants.\par
This approach especially allows the determination of the effective couplings $J_k$
as a function of the lattice parameters. They can also be derived
from the shape $\omega(q)$ of a softening phonon branch\cite{Iiz77}. A linear chain
with up to $N$th nearest neighbor couplings with dispersion relation
\begin{displaymath}
[\hbar \omega(q)]^2= \sum_{k=1}^N J_k\left(1-\cos kq\right)
\end{displaymath}
is considered as analogon to the crystal. A least-squares fit of this relation
to the soft phonon branch for various temperatures $T$ above the transition to
the modulated phases yields the dependence $J_k(T)$. 


\subsection{Structure predictions in terms of the SALM model and its derivatives}
For uniaxial modulations, the mean values of atomic displacements do not vary in planes
perpendicular to the axis of modulation, which is ${\vec c}$ in the
case of BCCD. Hence, the generalized variables have (at the thermal average) values
$Q^3_{\vec n}=Q_\ell(3)$, $Q^7_{\vec n}=Q_{\ell+1}(3)$ and 
$Q^4_{\vec n}=Q_\ell(4)$, $Q^8_{\vec n}=Q_{\ell+1}(4)$ where
$\ell=2n_3$ and $\ell+1=2n_3+1$ label consecutive layers (dotted vertical lines 
in figure \ref{figure:5})
formed from quadruples of
blocks belonging to cells at positions $(n_1n_2n_3)$. The displacements related to
the ,layer variables'
$Q_\ell(3)$ and $Q_\ell(4)$ are so-called ,layer modes'.
The notation is adapted to the fact that these layers can be considered
equivalent as the space group operation
${\op G}^5=\{ {\rm C}_2^z | \frac{1}{2} 1 \frac{1}{2} \}$
generates a pseudoperiodicity of half a lattice constant along the
${\vec c}$-axis by transforming one layer into the next. 
This construction
yields ,layer variables' as in Chen and Walker's 
model (cf.\ subsection \ref{subsection:CWmodel}). 
Specific differences arise for modulations along $\vec c$ (e.g.\ BCCD): The 
layers in Chen and Walker's model are located at positions $z_{\ell}=(2\ell+\frac{1}{4})$ and
$z_{\ell+1}=(2\ell+\frac{3}{4})$ and thus intersect 
(in our nomenclature) blocks belonging to different quadruples, as can 
be seen from figure \ref{figure:blocks} and figure \ref{figure:5} (solid vertical lines). 
Chen and Walker's choice of layer modes hence 
is a linear combination of the layer modes of the SALM model 
for modulations along $\vec c$. 
Two consecutive nitrogen atoms (along the $\vec c$-axis) are displaced simultaneously 
by each of the two modes for the respective layer they belong to.
For both models, the first layer mode displaces both atoms in the same direction along $\vec b$
whereas the $y$-displacements produced by the second layer mode are in opposite direction.
Depending on how one combines the nitrogen atoms and the layers, either both modes
are necessary for a satisfactory approximation of the displacements (Chen and Walker's model)
or only one (the layer mode related to $Q_\ell(3)$ in our nomenclature).
Hence, we tried  
a structure prediction for the displacements of the nitrogen atoms in the fourfold phase of 
BCCD either in terms of the 
layer modes of Chen and Walker's model or in terms of relevant cell modes.
If, following Chen and Walker, two consecutive nitrogen atoms are 
assigned to the planes given by full vertical lines in figure \ref{figure:5}, their displacements
are quite different and two modes are needed for a reasonable description. If, however,
the nitrogen atoms  are  assigned to the planes given by dotted lines, as proposed in the
present paper, their displacements  are very similar and one (relevant) mode 
gives already a very good description as shown in figure \ref{figure:5}.
An application of the DIS model to BCCD, 
which corresponds to taking into account a second set of SALMs
as opposed to the ANNNI model with one set of SALMs, yields even better agreement of
calculated and measured structures.\par
If both the ANNNI model and the DIS model are treated in mean
field approximation, there is no need in distinguishing pseudo spins
derived from amplitudes of cell modes for front or back half cells:
in an application of 
the ANNNI model to BCCD the pseudo spin $\sigma$ is derived from
$Q^3,Q^7$. In the DIS model the pseudo spin $\tau$ is derived 
from $Q^3,Q^7$ whereas the pseudo spin $\sigma$ is derived from $Q^4,Q^8$.\\
Since the pseudo spins $\sigma$ (for the ANNNI model) 
or $\sigma$ and $\tau$ (for the DIS model) stand for the amplitudes of
certain local modes, their transformation behavior under space group
operations can be determined and relations between the couplings can be found.
It can be shown\cite{Neu97} that 
the antisymmetric coupling $M$ between the two pseudo spin subsystems of the DIS
model is not introduced ad hoc but follows
necessarily from symmetry.\par
Making use of the transformation behavior of the generalized variables, expressions for the
spontaneous polarization of the crystal in terms of the amplitudes
$Q_\ell(3)$ and $Q_\ell(4)$ can be derived.
Since the components of the spontaneous polarization vector ${\vec P}_S$ 
must transform like any vectorial quantity under the operations of
the space group $Pnma$, only such terms are allowable in an expansion
of ${\vec P}_S$ in terms of relevant mode amplitudes that are in
accordance with symmetry. Expanding ${\vec P}_S$ up to second order
in $Q_\ell(3)$ and $Q_\ell(4)$ and keeping only terms involving same
or adjacent layers, the following expressions can be derived:
\begin{eqnarray*}
P_{S,x}&=& \sum_\ell  (-1)^\ell [P^1_x Q_\ell(3) Q_{\ell+1}(3) +  P^2_x Q_\ell(4) Q_{\ell+1}(4) + P^3_x Q_\ell(3) Q_\ell(4) \\[-1ex]
     & & \qquad\qquad + P^4_x \left( Q_\ell(3) Q_{\ell+1}(4) - Q_\ell(4) Q_{\ell+1}(3) \right)]
+ \ldots\\
P_{S,y}&=& \sum_\ell (-1)^\ell P^1_y Q_\ell(3) + \ldots, \\
P_{S,z}&=& \sum_\ell P^1_z Q_\ell(3) Q_\ell(4) + \ldots.
\end{eqnarray*}
$P^1_x,\ldots,P^4_x$, $P^1_y$ and $P^1_z$ are constants specific for
BCCD.
A sequence of equal
$Q_\ell$ leads to alternating signs of $x$- and $y$-displacements in
consecutive layers as the respective layer modes are transformed one into
another by the screw axis ${\op G}^5$. This is
as a consequence of the proposed definition of the local modes.
Thus, in the case of BCCD,
a correspondence of pseudo spin averages $S_\ell$ to $Q_\ell(3)$ 
with alternating signs was performed in our previous work\cite{Neu94a}.
The corresponding expressions for the DIS model with
\begin{eqnarray*}
Q_\ell(3) & \longrightarrow & (-1)^\ell t_\ell,\\
Q_\ell(4) & \longrightarrow & (-1)^\ell s_\ell
\end{eqnarray*}
read
\begin{eqnarray}
\label{equation:polariz}
P_{S,x}&=& \sum_\ell (-1)^\ell [ {\tilde P}^1_x t_\ell t_{\ell+1} +  {\tilde P}^2_x s_\ell s_{\ell+1} + {\tilde P}^3_x s_\ell t_\ell +{\tilde P}^4_x \left( t_\ell s_{\ell+1} - s_\ell t_{\ell+1} \right)], \nonumber \\
P_{S,y}&=& \sum_\ell {\tilde P}^1_y t_\ell, \nonumber \\
P_{S,z}&=& \sum_\ell {\tilde P}^1_z t_\ell s_\ell. 
\end{eqnarray}
For the values of pressure and temperature applied in measurements in BCCD\cite{Unr89},
the following spontaneous polarizations
are obtained depending on the wavenumber:~~
$\frac{m}{n} = \frac{\mbox{odd}}{\mbox{even}}$: $P_{S,x}$;~~
$\frac{m}{n} = \frac{\mbox{even}}{\mbox{odd}}$: $P_{S,y}$;~~
$\frac{m}{n} = \frac{\mbox{odd}}{\mbox{odd}}$: no spontaneous polarization.
This is in perfect agreement with the spontaneous polarizations obtained from the DIS model
{\em for the corresponding model parameters} and the assumption that a symmetry mode of
$\Lambda_3$-symmetry gives the primary contribution to the structural deformation.
If only one mode per half cell is retained
($Q^3,Q^7$), the expressions for $P_{S,x}$, $P_{S,y}$ and $P_{S,z}$ are given
by the terms in eqn.\ (\ref{equation:polariz}) depending only on $t_\ell$,
in complete accordance with the terms we proposed in ref.\cite{Neu94a}.
Recently new dielectric investigations on BCCD revealed the so-called
$T_S$-anomaly\cite{Sch96}
in the regions of the  commensurate $\frac{1}{4}$- and 
$\frac{1}{5}$-phases exihibiting the characteristics of a 
commensurate-commensurate phase transition where the wave number is unchanged.
The same phenomenon occurs in structures derived from the DIS model;
a detailed discussion will be given elsewhere\cite{Ple97b}.

\section{The influence of point defects}
\label{section:pointdefects}
In subsection \ref{subsection:theoderivedpd} it was shown that -- even if there
is not yet an ab initio type of theory -- it is possible to establish a 
quantitative connection
between experimentally determined $T,p$-phase diagrams and theoretical diagrams
formulated originally in terms of model parameters. In the case of an ANNNI model
description, the latter are the couplings $J_i$, defined, 
as explained above, by thermal averaging over local modes. A
simplifying step in this procedure was the plausible assumption that these model
parameters depend on $T$ and $p$ mainly via the total strains or via the lattice constants. By making
use of macroscopic material properties like thermal expansion, elastic 
coefficients etc., this lattice constant dependence of the couplings $J_i$ was
explicitely derived. The similarity of the
theoretical and the experimental phase diagrams support this procedure. The
success of this approach made it promising to interpret defect influences,
e.g.\ on phase diagrams, along the same lines. For the general treatment we shall
extend the basic assumption somewhat, allowing for a dependence of the effective
interactions not only on $T$ and dilatation but on $T$ and general
strain components. We shall first formulate this extended version, then include
defect contributions to the strains, and finally give some explicit results on
defect influences. The method is exemplified for the ANNNI model; it can, however,
easily be transposed to other models.\par
Let the phase diagram be formulated in the space of the experimental control
parameters temperature $T$ and components $\sigma^E_{ij}$ of the tensor
${\bfsigma}^E$ describing a homogenous external (applied) stress field.
The aim of the method to be described is to obtain an approximate theoretical
statement as to which one of the many modulated phases occuring in the model
will be stable at a point $(T,{\bfsigma}^E)$ in parameter space.\par
If results of ab initio electron theory or of atomistic interaction
potential theory followed by a thermal averaging procedure were at hand,
these would yield the effective interactions $J_i(T,{\bfsigma}^E)$ and the
stable phase could directly be read 
$(T,{\bfsigma}^E)$ from the theoretical ANNNI phase diagram in 
$(T,J_0,J_1,J_2)$-space for any set of control parameters.\par
If the dependence $J_i(T,{\bfsigma}^E)$ is not known from basic theory, the
following procedure, which makes use of macroscopic properties and characteristics
of the para phase boundary (PPB) yields approximate results: let the
empirically determined PPB and the modulation vectors on the PPB be given by
$T=T({\bfsigma}^E)$ and $q=q({\bfsigma}^E)$ and assume the effective model
parameters $J_i$ to depend on $T$ and ${\bfsigma}^E$ (only) via certain
linear combinations of the stress tensor components $\epsilon_{ij}$:
\begin{displaymath}
J_i= J_i(R^{(i)}_{mn}\epsilon_{mn})=: J_i(\Omega^{(i)})
\end{displaymath}
(summation over repeated subscripts). The coefficients $R^{(i)}_{mn}$ are chosen
in a plausible way taking the lattice structure into account. The
displacements are counted from the para phase positions at some temperature
$T_0$ above but close to the PPB, zero applied stresses ${\bfsigma}^E$ and
no defects. For crystals free of defects the total strains are, in linear
approximation, the sum of elastic and thermal expansion contributions:
\begin{equation}
\label{equation:formulax}
\epsilon_{mn}= S_{mnkj}\sigma^E_{kj}+\alpha_{mn}(T-T_0)
\end{equation}
and
\begin{equation}
\label{equation:formulay}
\Omega^{(i)}= R^{(i)}_{mn} S_{mnkj} \sigma^E_{kj} + R^{(i)}_{mn} \alpha_{mn}(T-T_0) =: \Lambda^{(i)}_{kj} \sigma^E_{kj} + B^{(i)}(T-T_0).
\end{equation}
For this step, values of the macroscopic elastic moduli and thermal expansion
coefficients are required. Relation (\ref{equation:formulay}) may be
(and was) improved by taking higher order terms and coefficients into account.
For any point (defined e.g.\ by the external stress ${\bfsigma}^E$) on the PPB,
the corresponding set of model parameters $J_i$ may be determined by 
observing that in the ANNNI model the modulation vector on the PPB depends only
on the interaction ratio $\kappa=-J_2/J_1$ [cf.\ eqn.\ (\ref{equation:qcrit})]:
\begin{displaymath}
q_{\text{crit}}(\kappa)= \frac{1}{2\pi} \arccos \frac{1}{4\kappa} \quad
\text{for }\kappa>0.25.
\end{displaymath}
Thus $\kappa$ is known for any point on the PPB. Now the model PPB is given by
\begin{displaymath}
kT_{\text{crit}}= 4J_0+(2\kappa+\frac{1}{4\kappa})J_1=: 4J_0 + f(\kappa) J_1.
\end{displaymath}
$J_0$ and $J_1$ can be determined from the (experimentally determined)
dependence of $T_{\text{crit}}$ on $f(\kappa)$ for variations of
${\bfsigma}^E$ which leave
\begin{displaymath}
\Omega^{(i)}=\Lambda^{(i)}_{kj} \sigma^E_{kj} + B^{(i)}(T_{\text{crit}}({\bfsigma}^E)-T_0),\,i=0,1
\end{displaymath}
and thus also $J_0$ and $J_1$ invariant. Once $\kappa,J_0,J_1$ are determined,
$J_2$ is obtained from $\kappa$ and $J_1$. In this way one obtains
corresponding values of $J_i$ and $\Omega^{(i)}$, that is a numerical
representation of the function $J_i(\Omega^{(i)})$. One can thus assign
to any point $(T,{\bfsigma}^E)$ (also to points off the PPB) the corresponding
values $J_i$. Reference to the ANNNI $(T,J_0,J_1,J_2)$-phase diagram
then yields the phase stable at $(T,{\bfsigma}^E)$.\par
For {\em defect crystals} the contributions $\bfepsilon^D$ of the point defects to the
local strains have to be included in eqn.\ (\ref{equation:formulax}).
The deformation due to a point defect (see e.g.\ ref.\cite{Kro58,Sie68}) is described
by its force dipole tensor $P_{ij}$. With the material's elastic Green's
function $G_{kj}({\vec r},{\vec r'})$, which is expected to be, in good
approximation, the same for different modulated phases, the strain field at
$\vec r$ produced by (equal) defects at positions ${\vec r}^\nu$ is
\begin{displaymath}
\epsilon^D_{mn}({\vec r})=\sum_\nu G_{mj,k^\nu n}({\vec r},{\vec r^\nu}) P_{kj}
\end{displaymath}
and that produced by an uncorrelated distribution of defects with local
defect density $\rho^D({\vec r})$ is
\begin{equation}
\label{equation:formulaxx}
\epsilon^D_{mn}({\vec r},\rho^D)=\int G_{mj,k'n}({\vec r},{\vec r'})  P_{kj} d{\vec r'}
\end{equation}
with $f_{,k'}:= \partial f/\partial x_{k'}$. In extension of the procedure
described above, the local model parameters, e.g.\ the ANNNI couplings
$J_i({\vec r})$ are then assumed to depend on the {\em local} strains
$\bfepsilon$ and via these\footnote{In addition, the interactions (between two units) will
be changed quite apart from the defect induced strain simply due to the fact that in one of 
the two units there is a substitutional or an interstitial atom. This effect is not
considered here.} not only on temperature and external stresses
$\bfsigma^E$ (e.g.\ on pressure), but by eqn.\ (\ref{equation:formulaxx})
also on the defect distribution $\rho^D({\vec r})$:
\begin{displaymath}
J_i({\vec r})=J_i\left(\bfepsilon({\vec r})\right)= J_i\left(\{S_{mnkj} \sigma^E_{kj}
+ \alpha_{mn}(T-T_0)+ \epsilon^D_{mn}({\vec r},\rho^D)\}\right).
\end{displaymath}
Dividing the defect strain into its average and a position dependent contribution
one has
\begin{displaymath}
\bfepsilon^D({\vec r})= \overline{\bfepsilon^D(\rho^D)}+ \left(
\bfepsilon^D({\vec r},\rho^D)-\overline{\bfepsilon^D(\rho^D)} \right).
\end{displaymath}
The influence of the {\em average strain} corresponds to an extra term
$\sigma^D_{kj}=C_{kjpq} \epsilon^D_{pq}$, which has to be added to the
external strain ${\sigma}^E_{kj}$. From the experimentally observed average
deformation $\tilde{\bfepsilon}$ connected with a defect concentration $\rho^D$,
the dipole tensor can be determined. One has
\begin{displaymath}
P_{kj}=C_{kjmn} \tilde{\epsilon}_{mn}/\rho^D.
\end{displaymath}
If our basic assumption is correct that the model parameters $J_i$ depend on
temperature and external stresses mainly via the total strain field including
elastic and thermal contributions and if the
main defect contribution is a function of the {\em average} defect induced strain
$\overline{\bfepsilon^D(\rho^D)}$,
then the dominant impurity effect on the phase diagram should be an alteration 
obtained by replacing the external stresses $\sigma^E_{kj}$ by
$\sigma^E_{kj}+\rho^D P_{kj}= \sigma^E_{kj}+C_{kjmn} \tilde{\epsilon}_{mn}$.
This relation was checked\cite{Rue94} for brominated BCCD under
hydrostatic pressure for which phase diagrams 
were experimentally determined\cite{Ao90,Mai92,Mai94} for various Br-concentrations.
Already the rough approximation of the variables $\Omega{(i)}$ by the
dilatation $\epsilon_{mn}/3$, i.e.\ the average relative volume change, leads
(with experimentally determined material constants\cite{Hau88,Hau89}) to
a good agreement with experimental data:
For Br-concentrations of 1.8\%;~7.3\%;~20\% the calculated pressure shifts are
-20 MPa;~-81 MPa;~-222 MPa as compared to experimentally observed shifts of
-18 MPa;~-61 MPa;~-238 MPa respectively.\par
The influence of {\em spatial variations of the defect strains} was investigated\cite{Rue94}
for a situation with simple model characteristics, control parameters and defect
arrangements: the coefficients $R$ in $\Omega^{(i)}$ were chosen to define
the dilatation part of the strains ($R^{(i)}_{mn}=\delta_{mn}$), the external
stress is a hydrostatic pressure ($\sigma^E_{ij}=-p\delta_{ij}$), and the
defect distribution is such as to produce a dilatation strain field
$\epsilon^D:=\epsilon^D_{ii}$ which varies only in one direction
(perpendicular to the modulation axis), with a position ($x$) dependence
described by a saw tooth function. The nearest neighbor interactions $J_0$ and
$J_1$ were set equal to each other. Depending on the amplitude of the spatial
variation of defect concentration or strain, this scenario leads to a coexistence
of several phases for given values of temperature and pressure. A small
section of the phase diagram is considered and the general procedure
described above for calculating the phase diagram is modified correspondingly:
the strain field and the free energy densities $f^\gamma$ of the phases $\gamma$
are expanded about values of temperature, pressure and position
corresponding now to a point in the middle of that section at a local position
in the middle of a saw tooth: 
\begin{displaymath}
f^\gamma\{J_i[\Omega(T,p,\epsilon^D(x))],T\}= f_0^\gamma+A^\gamma_T \Delta T 
+ A^\gamma_p \Delta p + A^\gamma_x \Delta x
\end{displaymath}
with $f^\gamma_0= f^\gamma(T^0,p^0,x^0)$ and
\begin{displaymath}
 A^\gamma_T:= \frac{\partial f^\gamma}{\partial J_i} \frac{\partial J_i}{\partial \Omega} \frac{\partial \Omega}{\partial T}+
\frac{\partial f^\gamma}{\partial T}; \quad A^\gamma_p:= \frac{\partial f^\gamma}{\partial J_i} \frac{\partial J_i}{\partial \Omega} \frac{\partial \Omega}{\partial p};\quad
A^\gamma_x:= \frac{\partial f^\gamma}{\partial J_i} \frac{\partial J_i}{\partial \Omega} \frac{\partial \Omega}{\partial \epsilon^D} \frac{\partial \epsilon^D}{\partial x}.
\end{displaymath}
With  the position $x^\gamma$ of the boundary between
phases $\gamma-1$ and $\gamma$, and with $\gamma=0$ the first and $\gamma=m$ the
last phase to occur, the total free energy is 
\begin{equation}
\label{equation:formulap}
F(p,T,\{x^\gamma\})=\sum_{\gamma=0}^m \int_{x^{\gamma-1}}^{x^\gamma} f^\gamma dx.
\end{equation}
Minimizing $F$ with respect to the boundary positions $x^\gamma$ yields the local
extensions of the different phases as function of $T$ and $p$.\par
Explicit calculations\cite{Rue94} were performed for the section of the
phase diagram show in figure \ref{figure:marc1}. The temperature (or
pressure) dependence of physical quantities which are different in different
phases (i.e.\ which depend on the modulation wave number $q$) will vary in
a typical way with the defect concentration gradient. As a representative of
such quantities we consider $q$ itself. Its average $\overline q$ is
obtained from the values $x^\gamma$, and the wave numbers $q^\gamma$.\par
Results of such calculations are shown in figure \ref{figure:marc2}. For
crystals with a spatially constant defect density, ${\overline q}(T)$ 
is characterized by a series of
plateaus corresponding to one modulated phase each with discontinuous
transitions between them (figure \ref{figure:marc2}a). Upon introducing
a spatial (saw-tooth-like) variation of the defect density,
the plateaus become gradually narrower
and the transitions become continuous.
This is shown in figure \ref{figure:marc2}: the amplitude of the defect
density variation increases in equal steps from figure \ref{figure:marc2}a
(amplitude zero) to figure \ref{figure:marc2}f.
The transitions intervals now correspond to the coexistence of two or more
phases whose boundaries are continuously shifted with temperature. For
sufficiently high concentration gradients some plateaus disappear completely,
indicating that the stability range of the corresponding phase $\gamma$ is so small
that there is no temperature for which $\gamma$ is the only stable phase in the whole
specimen.

%
\section{Discussion and outlook}
\label{section:discoutlook}
The theoretical description of materials exhibiting modulated phases may be subdivided into 
four main points:\par
1) Eventually the description should be based on an ab initio electron theory (e.g.\ ref\cite{Kat86}) or, if that is too
difficult to handle with the necessary accuracy for the rather large number of atoms per unit cell,
on calculations using empirical atomic interactions (e.g.\ ref\cite{Etx92}). A suitable procedure would be to determine total
energies for the para phase configuration and for a
number of appropriately chosen deformations. The respective results should allow the selection of adequate 
simplified models, the formulation of the respective effective Hamiltonians, and a determination
of the model parameters. These models should then be handled as described under points 2),3),4). So
far there are scarcely any results of this kind. 
Calculations along these lines, however, would allow to go, for a not too complicated 
system, all the way from  an atomistic theory to explicit statements 
on phase diagrams, polarizations etc.
In such a treatment
approximations are of course necessary and should be made (in the original atomistic theory as well as in 
the consecutive model calculations); but one should avoid ad hoc assumptions (fits) as far as possible.
\par
2) 
The second part of the description is the formulation and investigation of the properties of model systems
with one or a few relevant local variables like e.g.\ the DIFFOUR model. Especially if this is the
first step (i.e.\ if one cannot deduce these models explicitely from one of the ab initio theories of
point 1), all symmetry information should be taken into consideration in the formulation of the models.
The main task is the statistical mechanics treatment which should start from the respective (effective)
Hamiltonian and should yield information on phase diagrams, orders of the transitions, polarizations
etc. The calculations may be and have been performed either in a more or less analytic way by using schemes
of approximation like e.g.\ the mean field or the self consistent phonon method. Or, alternatively,
numerical approaches (molecular dynamics simulations) were used. So far, calculations for such models
were performed mainly for very simplified geometries like linear chains.\par
3)
A further step in the simplification process is the projection of models with a (small) number of
continuous local variables onto models with discrete states, or, finally, onto pseudo spin models.
Depending on the number of relevant original local modes one obtains pseudo spins with one or more
components per crystallographic unit. Substantial advantages of this type of description are,
that the powerful apparatus of Monte Carlo simulations can be applied and
that analytic approximations exist which allow the explicit calculation of many interesting physical
properties.
This part of
the theory is well developed. This holds especially for mean field treatments which in some instances
where compared with more pretentious methods and shown to yield a good description.\par
4) Finally there is the problem of translating the results of model calculations obtained as functions
of the model parameters into data which can be quantitatively compared to experimental data, i.e.\ which
are expressed as functions of experimentally given quantities like pressure or stresses (besides
temperature). In section \ref{section:application} it was shown how this can be done. By making use of
macroscopic material constants, a physically motivated mapping from one parameter set to the other
can be performed. For the material considered (BCCD) the results show a satisfactory agreement with experimental data.\par
In the last years considerable effort has been addressed to the approaches described in section 
\ref{section:PSmodels}, that is to the formulation of various $p$-state and pseudo spin models and
to their statistical mechanics. Methods applied were MFA calculations, exact series expansions for
selected regions in parameter space, and Monte Carlo simulations. Combining these approaches, a
reliable theoretical determination of most of the interesting thermodynamic properties of such models is
possible.\par
A transformation of these model properties into a form directly comparable to experimental data can be
performed by the procedure described in subsection \ref{subsection:theoderivedpd} or by a similar
scheme. An interesting by-product of this method is information on the dependence of the model
parameters on the experimentally given state variables and especially on the lattice distortions.
Combined with a consideration of experimentally determined atomic positions in some of the more stable
phases, this should further the understanding of the local dynamics responsible for the transition sequences.\par
Of special interest, finally, would be the determination of model parameters -- for not too complicated
materials -- from an atomistic theory. It should be possible, with the nowadays available computing power,
to carry through the necessary calculations. Investigations of this type would clarify the details of
the relevant atomistic mechanisms and e.g.\ help to understand why the phase diagrams of members of the
same homologous family of materials may differ drastically.

%
\section{Acknowledgements}
Interesting discussions with Prof.\ D.\ Sannikov and financial support from the Deutsche
Forschungsgemeinschaft (Si 358/1) are gratefully acknowledged.

\clearpage

\begin{figure}
\caption{The blocks ${\frak A}^i_{\vec n},\; i=1\ldots8$ belonging to the unit cell
(depicted as wireframe)
at the origin {\bf O}. \label{figure:blocks}}
\end{figure}

\begin{figure}
\caption{Mean field phase diagram of the ANNNI model with $J_0=J_1$.
\label{figure:annnimeanfield}}
\end{figure}

\begin{figure}
\caption{The different fixed point types. Shown are the positions of the four
eigenvalues of the matrix $\underline{\underline{G}}^k$ in the complex plane.}
\label{figure:fixpoint}
\end{figure}

\begin{figure}
\caption{Ground state of the DIS model.
The pair of upper/lower signs per column of the symbols represent
the signs of the $\tau$- and $\sigma$-spin averages in consecutive layers:
I is the fully ferroelectric phase, II the
fully antiferroelectric phase, whereas III and IV are mixed phases in which one
spin type orders ferroelectrically, the other antiferroelectrically.
The structure of the phase V repeats itself after 4 layers.
The vertices are given by ($\frac{L}{M}$,
$\frac{K}{M}$) = (1,0), (0,1), (-1,0) and (0,-1).}
\label{figure:disground}
\end{figure}

\begin{figure}
\caption{Global phase diagram of the DIS model with $\kappa_- = - 0.1$ and
$\frac{J}{M} = \frac{J'}{M} = -0.25$.
Only few modulated phases are shown.
$\theta = \frac{k_BT}{M}$ is the reduced temperature, 
$\kappa_\pm = \frac{K \pm L}{M}$, and $K$, $L$, $M$ are
the different couplings in direction of the modulations.}
\label{figure:diagram1}
\end{figure}

\begin{figure}
\caption{Global phase diagram of the DIS model with $\lambda=\frac{L}{M} = 0.1$ and
$\frac{J}{M} = \frac{J'}{M} = -0.25$.}
\label{figure:diagram2}
\end{figure}

\begin{figure}
 \caption{\label{figure:ANNNIpTPD} $(p,T)$-phase diagram for BCCD calculated from 
 the ANNNI model.  Dashed lines: boundaries to higher commensurate or incommensurate phases.}
 \end{figure}

\begin{figure}
\caption{Measured ($\circ$) and calculated ($\bullet$)
$y$-displacements of the nitrogen atoms in the fourfold phase of BCCD.
The modulation profile was calculated from the ANNNI
model, i.e.\ with one variable per unit (crystallographic half cell). Solid and dotted 
lines: centers of layers in Chen and Walker's model and
in the SALM model respectively; consecutive
nitrogen atoms are assigned to the layer between them.
\label{figure:5}}
\end{figure}

\begin{figure}
\caption{Section of the ANNNI model phase diagram exhibiting some of the more stable phases for which the results shown in figure \protect\ref{figure:marc2} were obtained. This section is part of the region occupied by the phases observed in BCCD.\label{figure:marc1}}
\end{figure}

\begin{figure}
\caption{Averaged modulation wave number $q$ vs.\ $\Theta=kT/J_1$. Plateaus correspond to
phases $\langle 56\rangle$
(1), $\langle (56)^2556\rangle$ (2), $\langle 56556 \rangle$ (3),
$\langle 56(556)^2 \rangle$ (4) and $\langle 556 \rangle$ (5). $\rho^D_{,x}$ and $a_{,x}$
are increasing from a) ($a=\mbox{const}$) to f) in equal steps. In all six cases
the sum of applied pressure and chemical
shift at $x_0$ is kept constant (e.g.\ by keeping $\rho^D(x_0)$ and
$a(x_0)$ fixed).\label{figure:marc2}}
\end{figure}

\begin{thebibliography}{100}

\bibitem{Bli86a}
R.~Blinc and A.P. Levanyuk, editors.
\newblock {\em Incommensurate Phases in Dielectrics, 1.\ Fundamentals}.
\newblock Amsterdam, 1986.

\bibitem{Sel92}
W.~Selke.
\newblock Spatially modulated structures in systems with competing
interactions.
\newblock In C.~Domb and J.L. Lebowitz, editors, {\em Phase Transitions and Critical Phenomena}, volume~15, pages 1--72. Academic Press, 1992.


\bibitem{Cum90}
H.Z. Cummins.
\newblock {\em Physics Reports}, 195:211--409, 1990.

\bibitem{Saw92}
S.~Sawada, M.~Takashige, T.~Yamaguchi, F.~Shimizu, and H.~Suzuki.
\newblock {\em Ferroelectrics}, 137:205, 1992.

\bibitem{Per89}
J.M. P\'{e}rez-Mato, F.J. Z\'{u}\~{n}iga, and G.~Madariaga.
\newblock {\em Phase Transitions}, 16/17:439--444, 1989.

\bibitem{Bli86b}
R.~Blinc and A.P. Levanyuk, editors.
\newblock {\em Incommensurate Phases in Dielectrics, 2.\ Materials}.
\newblock Amsterdam, 1986.

\bibitem{Sai91}
P.~Saint-Gr{\'e}goire, R.~Almairac, A.~Astito, J.~Lapasset, and J.~Moret.
\newblock {\em Solid State Comm.}, 80:451, 1991.

\bibitem{Axe86}
J.D. Axe, M.~Iizumi, and D.~Shirane.
\newblock Phase transformations in \chemical{K_2SeO_4} and structurally related
  insulators.
\newblock In R.~Blinc and A.P. Levanyuk, editors, {\em Incommensurate Phases in
  Dielectrics, 2.\ Materials}. Amsterdam, 1986.

\bibitem{Ten90b}
Th.~Tentrup and R.~Siems.
\newblock {\em Ferroelectrics}, 105:813, 1990.

\bibitem{Neu94b}
B.~Neubert.
\newblock Diploma thesis, Saarbr\"ucken, 1994.

\bibitem{Jan86a}
T.~Janssen.
\newblock Microscopic theories of incommensurate crystal phases.
\newblock In R.~Blinc and A.P. Levanyuk, editors, {\em Incommensurate Phases in
  Dielectrics, 1.\ Fundamentals}. Amsterdam, 1986.


\bibitem{Jan86b}
T.~Janssen.
\newblock {\em Ferroelectrics}, 66:203--216, 1986.

\bibitem{Jan91}
T.~Janssen.
\newblock {\em Ferroelectrics}, 124:41, 1991.

\bibitem{Sel88}
W.~Selke.
\newblock {\em Physics Reports}, 170:213, 1988.

\bibitem{Yeo88}
J.~Yeomans.
\newblock {\em Solid State Physics}, 41:151, 1988.

\bibitem{Ple94}
M.~Pleimling and R.~Siems.
\newblock {\em Ferroelectrics}, 151:69--74, 1994.

\bibitem{Ple96}
M.~Pleimling and R.~Siems.
\newblock {\em Ferroelectrics}, 185:103, 1996.

\bibitem{Kur94}
M.~Kurzy{\'n}ski.
\newblock {\em Phase Transitions}, 52:1, 1994.

\bibitem{Che90}
Z.Y. Chen and M.B. Walker.
\newblock {\em Phys.\ Rev.\ Lett.}, 65:1223, 1990.

\bibitem{Che91b}
Z.Y. Chen and M.B. Walker.
\newblock {\em Phys. Rev.}, B43:5634, 1991.

\bibitem{Tol87}
J.-C Tol\'edano and P.~Tol\'edano.
\newblock {\em The Landau Theory of Phase Transitions}.
\newblock World Scientific, 1987.

\bibitem{Koc90}
J.~Koci\'{n}ski.
\newblock {\em Commensurate and Incommensurate Phase Transitions}.
\newblock Elsevier, Amsterdam, 1990.

\bibitem{San89b}
D.G. Sannikov.
\newblock {\em Sov. Phys. JETP}, 69:1244, 1990.

\bibitem{San90b}
D.G. Sannikov.
\newblock {\em Sov. Phys. JETP}, 70:1144, 1990.

\bibitem{San91}
D.G. Sannikov.
\newblock {\em Ferroelectrics}, 124:49, 1991.

\bibitem{San97}
D.G. Sannikov and G.~Schaack.
\newblock {\em preprint}, 1996.

\bibitem{Rib90}
J.L. Ribeiro, J.C. Tol\'edano, M.R. Chaves, A.~Almeida, H.E. M\"user,
  J.~Albers, and A.~Kl\"opperpieper.
\newblock {\em Phys.\ Rev.\ B}, 41:2343, 1990.

\bibitem{Alm92}
A.~Almeida, M.R. Chaves, J.M. Kiat, J.~Schneck, W.~Schwarz, J.-C. Tol\'edano,
  J.L. Ribeiro, A.~Kl\"opperpieper, H.E. M\"user, and J.~Albers.
\newblock {\em Phys. Rev.}, B45:9576, 1992.

\bibitem{Cha93}
M.R. Chaves, A.~Almeida, J.-C. Tol\'edano, J.~Schneck, J.M. Kiat, W.~Schwarz,
  J.L. Ribeiro, A.~Kl\"opperpieper, J.~Albers, and H.E. M\"user.
\newblock {\em Phys. Rev.}, B48:13318, 1993.

\bibitem{Neu97}
B.~Neubert, M.~Pleimling, and R.~Siems.
\newblock Symmetry-based discrete lattice models for structurally modulated
  materials.
\newblock {\em to be published}.

\bibitem{Tho71}
H.~Thomas.
\newblock Structural phase transitions and soft modes.
\newblock In {\em Proceedings of the NATO Advanced Study Institute}. Oslo,
  1971.

\bibitem{Neu94a}
B.~Neubert, M.~Pleimling, Th. Tentrup, and R.~Siems.
\newblock {\em Ferroelectrics}, 155:359--364, 1994.

\bibitem{Neu96}
B.~Neubert and R.~Siems.
\newblock From crystal structure to phase diagrams: Use of {ANNNI}- and related
  models.
\newblock {\em Ferroelectrics}, 185:95--98, 1996.

\bibitem{Hli96}
J.~Hlinka, O.~Hernandez, M.~Quilichini, and R.~Currat.
\newblock {\em Ferroelectrics}, 185:221--224, 1996.

\bibitem{Qui96}
M.~Quilichini and J.~Hlinka.
\newblock {\em Ferroelectrics}, 183:215--224, 1995.


\bibitem{Per88}
J.M. P\'{e}rez-Mato.
\newblock {\em Solid State Comm.}, 67:1145--1150, 1988.

\bibitem{Ezp92}
J.M. Ezpeleta, F.J. Z\'{u}\~{n}iga, J.M. P\'{e}rez-Mato, W.A. Paciorek, and
  T.~Breczewski.
\newblock {\em Acta Cryst.}, B48:261, 1992.

\bibitem{Ell71}
R.J. Elliot.
\newblock The Pseudo-Spin Formalism For Displacive Transitions.
\newblock In {\em Proceedings of the NATO Advanced Study Institute}. Oslo, 1971.

\bibitem{Gil74}
N.S. Gillis and T.R. Koehler.
\newblock {\em Phys.\ Rev.\ B}, 9:3806, 1974.

\bibitem{Sta76b}
S.~Stamenkovi{\'{c}}, N.M. Plakida, V.L. Aksienov, and T.~Sikl{\'{o}}s.
\newblock {\em Phys.\ Rev.\ B}, 14:5080, 1976.

\bibitem{Kra92}
M.~Krauzman, M.~Debeau, R.M. Pick, M.~Quilichini, P.~Launois, and F.~Moussa.
\newblock {\em J. Phys. I France}, 2:329, 1992.

\bibitem{Deh29}
U.~Dehlinger.
\newblock {\em Annalen der Physik 5.~Folge}, 2:749--793, 1929.

\bibitem{Fre38}
Y.I. Frenkel and T.~Kontorowa.
\newblock {\em Zh. Eksp. Teor. Fiz.}, 8:1340, 1938.

\bibitem{Fra49}
F.C. Frank and J.H. Van~der Merwe.
\newblock {\em Proc. R. Soc.}, A198:205, 1949.

\bibitem{Jan81a}
T.~Janssen and J.A. Tjon.
\newblock {\em Phys. Rev. B}, 24:2245--2258, 1981.

\bibitem{Jan82}
T.~Janssen and J.A. Tjon.
\newblock {\em Phys. Rev. B}, 25:3767--3785, 1982.

\bibitem{Jan83c}
T.~Janssen and J.A. Tjon.
\newblock {\em J. Phys. C}, 16:4789--4810, 1983.

\bibitem{Che91a}
Z.Y. Chen and M.B. Walker.
\newblock {\em Phys.\ Rev.}, B43:760, 1991.

\bibitem{Fol91}
I.~Folkins, Z.Y. Chen, and M.B. Walker.
\newblock {\em Phys.\ Rev.}, B44:374, 1991.

\bibitem{Kap93}
C.~Kappler and M.B. Walker.
\newblock {\em Phys.\ Rev.}, B48:5902, 1993.

\bibitem{Ell61}
R.J. Elliott.
\newblock {\em Phys.\ Rev.}, 124:346, 1961.

\bibitem{Sel85}
W.~Selke, M.~Barreto, and J.~Yeomans.
\newblock {\em J. Phys. C: Solid State Phys.}, 18:L393, 1985.

\bibitem{Bar85}
M.~Barreto and J.~Yeomans.
\newblock {\em Physica}, 134A:84, 1985.

\bibitem{Ran85}
J.~Randa.
\newblock {\em Phys. Rev. B}, 32:413, 1985.

\bibitem{Vil80}
J.~Villain and M.~Gordon.
\newblock {\em J. Phys.}, C13:3117, 1980.

\bibitem{Mas83}
H.~Mashiyama.
\newblock {\em J. Phys. C: Solid State Phys.}, 16:187--201, 1983.

\bibitem{Bak80}
P.~Bak and J.~von Boehm.
\newblock {\em Physical Review}, B21, 1980.

\bibitem{Sel84b}
W.~Selke and P.M. Duxbury.
\newblock {\em Z. Phys.}, B57:49, 1984.

\bibitem{Dux83}
P.M. Duxbury and W.~Selke.
\newblock {\em J.\ Phys.\ A: Math.\ Gen.}, 16:L741, 1983.

\bibitem{Sie89a}
R.~Siems and Th. Tentrup.
\newblock {\em Phase Transitions}, 16/17:287, 1989.

\bibitem{Ten90a}
Th. Tentrup and R.~Siems.
\newblock {\em Ferroelectrics}, 105:385, 1990.

\bibitem{Ten92}
Th. Tentrup and R.~Siems.
\newblock {\em Ferroelectrics}, 125:75, 1992.

\bibitem{Jen94a}
K.~Jenal, T.~Tentrup, and R.~Siems.
\newblock {\em Ferroelectrics}, 155:353--358, 1994.

\bibitem{Jen96}
K.\ Jenal and R.\ Siems.
\newblock {\em Ferroelectrics}, 185:99, 1996.

\bibitem{Yok81}
C.S.O. Yokoi, M.P. Coutinho-Filho, and S.R. Salinas.
\newblock {\em Phys. Rev.}, B24:4047, 1981.

\bibitem{Nak89}
K.~Nakanishi.
\newblock {\em J. Phys. Soc. Jpn.}, 58:1296, 1989.

\bibitem{Yok91}
C.S.O. Yokoi.
\newblock {\em Phys. Rev.}, B43:8487, 1991.

\bibitem{Nak92}
K.~Nakanishi.
\newblock {\em J. Phys. Soc. Jpn.}, 61:2901, 1992.

\bibitem{Rot93}
F.~Rotthaus and W.~Selke.
\newblock {\em J. Phys. Soc. Jpn.}, 62:378, 1993.

\bibitem{Axe81}
F.~Axel and S.~Aubry.
\newblock {\em J.\ Phys.}, C14:5433, 1981.

\bibitem{Jan83a}
T.~Janssen and J.A. Tjon.
\newblock {\em J.\ Phys.}, A16:673, 1983.

\bibitem{Hog84}
M.~H{\o}gh~Jensen and P.~Bak.
\newblock {\em Phys. Rev. B}, 29:6280, 1984.

\bibitem{Ten88b}
Th. Tentrup and R.~Siems.
\newblock {\em Z. Phys. B. - Condensed Matter}, 72:503, 1988.

\bibitem{Sie89b}
R.~Siems and Th. Tentrup.
\newblock {\em Ferroelectrics}, 98:303, 1989.

\bibitem{Fis80}
M.E. Fisher and W.~Selke.
\newblock {\em Phys.\ Rev.\ Lett.}, 44:1502, 1980.

\bibitem{Fis81}
M.E. Fisher and W.~Selke.
\newblock {\em Philos.\ Trans.\ Royal Soc.}, 302:1, 1981.

\bibitem{Fis87b}
M.E. Fisher and A.M. Szpilka.
\newblock {\em Phys. Rev.}, B36:5343, 1987.

\bibitem{Yeo84}
J.~M. Yeomans and M.~E. Fisher.
\newblock {\em Physica}, A127:1, 1984.

\bibitem{Yeo82}
J.~M. Yeomans.
\newblock {\em J. Phys. C: Solid State Phys.}, 15:7305, 1982.

\bibitem{Sen93}
F.~Seno, D.A. Rabson, and J.M. Yeomans.
\newblock {\em J. Phys. A: Math. Gen.}, 26:4887--4905, 1993.

\bibitem{Red77a}
S.~Redner and H.E. Stanley.
\newblock {\em J. Phys.}, C10:4765, 1977.

\bibitem{Red77b}
S.~Redner and H.E. Stanley.
\newblock {\em Phys. Rev.}, B16:4901, 1977.

\bibitem{Oit85}
J.~Oitmaa.
\newblock {\em J. Phys.}, A18:365--375, 1985.

\bibitem{Mo91}
Z.~Mo and M.~Ferer.
\newblock {\em Phys. Rev.}, B43:10890--10905, 1991.

\bibitem{Gar76}
T.~Garel and P.~Pfeuty.
\newblock {\em J. Phys.}, C9:L245, 1976.

\bibitem{Dro76}
M.~Droz and M.D. Coutinho-Filho.
\newblock Critical behaviour of magnetic systems with helical state.
\newblock In {\em Magnetism and Magnetic Materials--1975 (21st Annual
  Conference---Philadelphia)}, page 465, 1976.

\bibitem{Kur86}
M.~Kurzy{\'n}ski and M.~Halawa.
\newblock {\em Phys. Rev.}, B34:4846, 1986.

\bibitem{Kur90}
M.~Kurzy{\'n}ski, M.~Ossowski, and M.~Bartkowiak.
\newblock {\em Ferroelectrics}, 105:107, 1990.

\bibitem{Kur92b}
M.~Kurzy{\'n}ski and M.~Bartkowiak.
\newblock {\em J. Phys.: Condensed Matter}, 4:2609--2614, 1992.

\bibitem{Ple97b}
M.~Pleimling, B.~Neubert, and R.~Siems.
\newblock to be published.

\bibitem{Ple97a}
M.~Pleimling, B.~Neubert, and R.~Siems.
\newblock to be published.

\bibitem{Sch96a}
P.D. Scholten and D.R. King.
\newblock {\em Phys. Rev.}, B53:3359, 1996.

\bibitem{Ao89}
R.~Ao, G.~Schaack, M.~Schmitt, and M.~Z\"oller.
\newblock {\em Phys. Rev. Letters}, 62:183a, 1989.

\bibitem{Kir93}
B.~Kirchner, M.~Le~Maire, G.~Schaack, and M.~Schmitt-Lewen.
\newblock {\em Europhys.\ Lett.}, 22:113, 1993.

\bibitem{Iiz77}
M.~Iizumi, J.D. Axe, and G.~Shirane.
\newblock {\em Phys.\ Rev. B}, 15:4392, 1977.

\bibitem{Unr89}
H.-G. Unruh, F.~Hero, and V.~Dvo\v{r}\'{a}k.
\newblock {\em Solid State Commun.}, 1989.

\bibitem{Sch96}
G.~Schaack, M.~Le~Maire, M.~Schmitt-Lewen, M.~Illing, A.~Lengel, M.~Manger, and
  R.~Straub.
\newblock {\em Ferroelectrics}, 183:205--214, 1996.

\bibitem{Kro58}
E.~Kr\"{o}ner.
\newblock {\em Kontinuumstheorie der Versetzungen und Eigenspannungen}.
\newblock Berlin, 1958.

\bibitem{Sie68}
R.~Siems.
\newblock Wechselwirkung von Defekten in Kristallen, Berichte der Kernforschungsanlage J\"{u}lich, Nr.\ 545, 1968.

\bibitem{Rue94}
M.~R{\"u}hl{\"a}nder and R.~Siems.
\newblock {\em Ferroelectrics}, 155:365--370, 1994.

\bibitem{Ao90}
R.~Ao, G.~Lingg, G.~Schaack, and M.~Z\"oller.
\newblock {\em Ferroelectrics}, 105:391--396, 1990.

\bibitem{Mai92}
M.~Le~Maire, G.~Lingg, G.~Schaack, M.~Schmitt-Lewen, G.~Strau\3, and
  A.~Kl\"opperpieper.
\newblock {\em Ferroelectrics}, 125:87--92, 1992.

\bibitem{Mai94}
M.~Le~Maire, A.~L\'{o}pez~Ayala, G.~Schaack, A.~Kl\"opperpieper, and H.~Metz.
\newblock {\em IMF8}, 1994.

\bibitem{Hau88}
S.~Hauss\"uhl, J.~Liedtke, J.~Albers, and A.~Kl\"opperpieper.
\newblock {\em Z.\ Phys.}, B70:219, 1988.

\bibitem{Hau89}
S.~Hauss\"uhl.
\newblock private communication, 1989.

\bibitem{Kat86}
V.~Katkanant, P.J. Edwardson, J.R. Hardy, and L.L. Boyer.
\newblock {\em Phys. Rev. Lett.}, 57:2033, 1986.

\bibitem{Etx92}
I.~Etxebarria, J.M. Perez-Mato, and G.~Madariaga.
\newblock {\em Phys.\ Rev.\ B}, 46:2764, 1992.


\end{thebibliography}
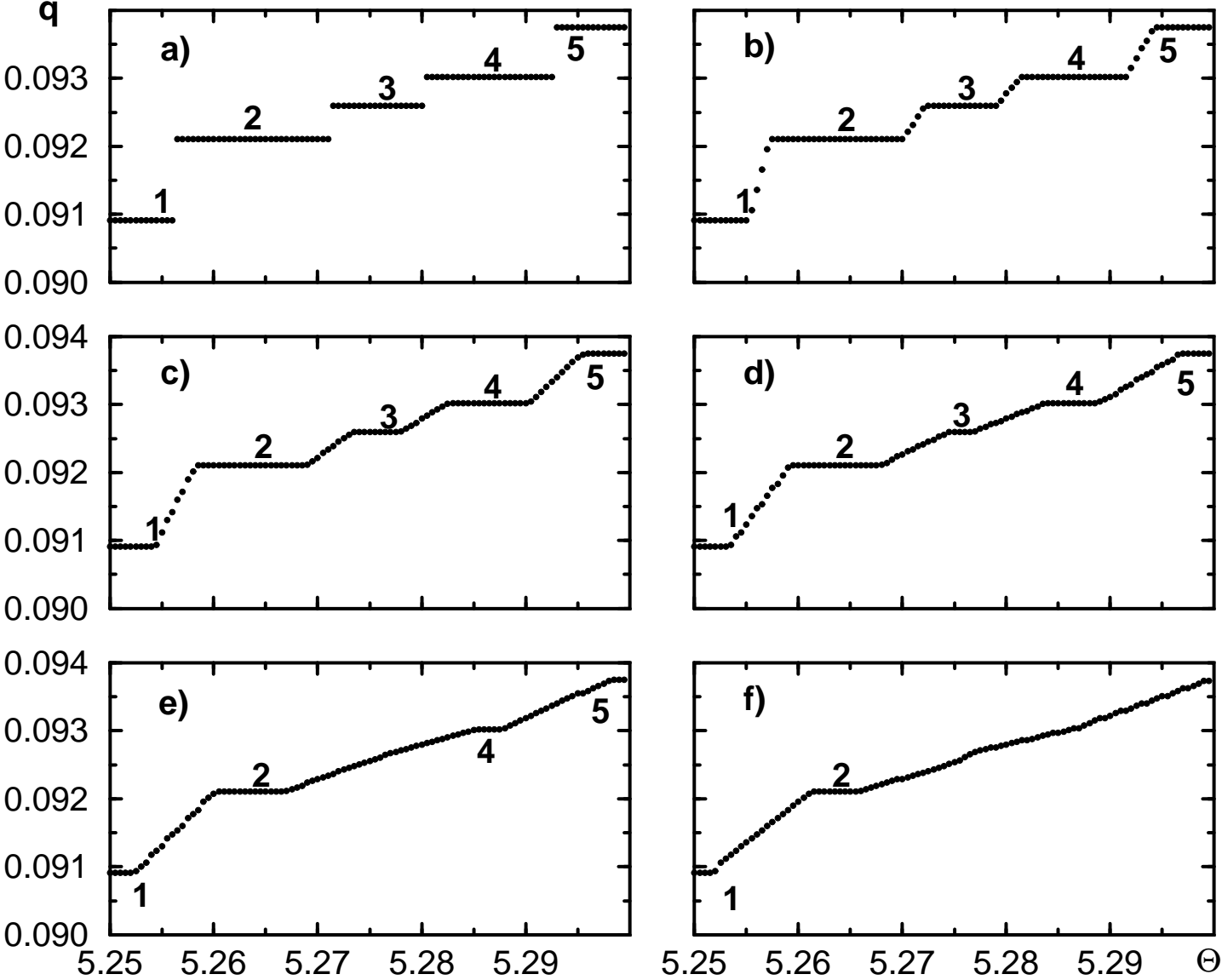
\end{document}